%% file: Paper.tex
\documentclass[11pt]{article}

\usepackage[utf8]{inputenc}
\usepackage[T1]{fontenc}
\usepackage[english]{babel}

\usepackage{amsmath,amssymb,amsthm}

\usepackage{graphicx}
\usepackage{subcaption}
\usepackage{booktabs}
\usepackage[table]{xcolor}
\usepackage{float}
\usepackage{placeins}

\usepackage[letterpaper,margin=1in]{geometry}

\usepackage{natbib}

\usepackage[hidelinks]{hyperref}

\theoremstyle{plain}
\newtheorem{theorem}{Theorem}[section]

\theoremstyle{definition}

\title{A Proper Random Walk Spline Model for Bayesian Smoothing}

\author{
Eman Kabbas$^{1}$, Janet van Niekerk$^{2}$, H{\aa}vard Rue$^{3}$\\[0.8em]
\small $^{1}$Jubail Industrial College,\\
\small Royal Commission for Jubail and Yanbu, Jubail, Saudi Arabia\\[0.4em]
\small $^{2}$Department of Statistics, University of Pretoria,\\
\small Pretoria, South Africa\\[0.4em]
\small $^{3}$Statistics Program, CEMSE Division,\\
\small King Abdullah University of Science and Technology (KAUST),\\
\small Thuwal, Saudi Arabia
}

\date{}

\begin{document}
\maketitle

\begin{abstract}
We introduce the \emph{Proper Random Walk} (PRW) family of spline models with a particular focus on the \emph{Second-Order Proper Random Walk} (PRW2), a proper stationary alternative to the standard random walk of order 2 (RW2) spline.
The PRW2 retains the computational strengths and spline-like smoothing of the RW2 while providing uniform marginal variance and stable uncertainty behavior. The PRW2 model is governed by a range parameter that controls correlation decay: in the infinite-range limit, it recovers RW2. In addition to its discrete representation, we show that the PRW2 model admits a continuous-time formulation as the critically damped Mat\'ern model with smoothness $\nu=3/2$, which provides an interpretable parameterization and enables construction through a Galerkin finite-element approximation on general grids. To mitigate edge distortions on bounded domains, we propose a local boundary correction that enforces the stationary joint law while preserving sparsity and $O(n)$ factorization. Finally, we develop penalized complexity priors for the model hyperparameters, enabling principled shrinkage toward the RW2 limit. The model is applied to simulated and real data to illustrate its properties and advantages.

\end{abstract}

\thispagestyle{empty}

\section{Introduction}

\subsection{Background}
Splines provide a flexible and widely used framework for representing unknown smooth functions while controlling roughness through regularization. Classical regression splines approximate a function using a fixed set of basis functions and estimate the coefficients by least squares, while smoothing splines add a penalty on curvature to balance fidelity and smoothness. Modern spline variants such as $B$-splines and $P$-splines further improve numerical stability and modeling flexibility by combining structured bases with explicit penalization. From a Bayesian perspective, many of these approaches can be expressed through Gaussian priors on basis coefficients, which yields coherent uncertainty quantification and convenient embedding within hierarchical models. This probabilistic view also clarifies the link between spline penalties and Gaussian Markov random fields (GMRFs), enabling scalable inference via sparse precision matrices.

Among Bayesian spline models, the random walk of order 2 (RW2) model is a standard choice for smoothing due to its simplicity, physical interpretation, computational efficiency, and wide use in large-scale Bayesian smoothing and prediction. The RW2 model is commonly used as a latent component to represent unknown smooth effects, including non-linear covariate responses and structured temporal trends within hierarchical models. \citet{lezama2020using} used RW2 smoothers to capture flexible predictor--response relationships in ecological and fisheries applications fitted with the integrated nested Laplace approximation and stochastic partial differential equation (INLA--SPDE) approach, and \citet{konstantinoudis2023workflow} applied RW2 to model non-linear temperature effects in high-resolution excess-mortality estimation. For temporal structure, \citet{johnson2024novel} used RW2 to capture serial dependence in environmental time series such as weekly PM$_{2.5}$ concentrations. The same second-order trend recurs throughout spatio-temporal public-health modeling --- in child mortality estimation \citep{amegbor2023spatiotemporal,li2019changes}, dengue risk mapping \citep{salim2025predicting}, and infectious-disease modeling with alternative spatial priors \citep{subramanian2024modelling} --- as well as in climate applications, where it serves as a long-term trend component in modeling fire occurrence \citep{valente2023spatio}.

Despite its practical success, the standard RW2 model used in latent Gaussian models is \emph{intrinsic}: its precision matrix is rank deficient and does not define a proper joint distribution without additional identifiability constraints (e.g., sum-to-zero or orthogonality to low-order polynomials). These constraints are often introduced implicitly by software defaults, but they are not innocuous: they act globally, and therefore shape the uncertainty reported by the model. In what follows, we characterize several properties implied by the intrinsic formulation. These are inherent consequences of the model definition rather than artifacts of implementation.

\subsection{The variance trap of RW2 as an intrinsic spline model}

\begin{figure}[ht]
  \centering

  \begin{subfigure}[t]{0.32\textwidth}
    \centering
    \includegraphics[width=\linewidth]{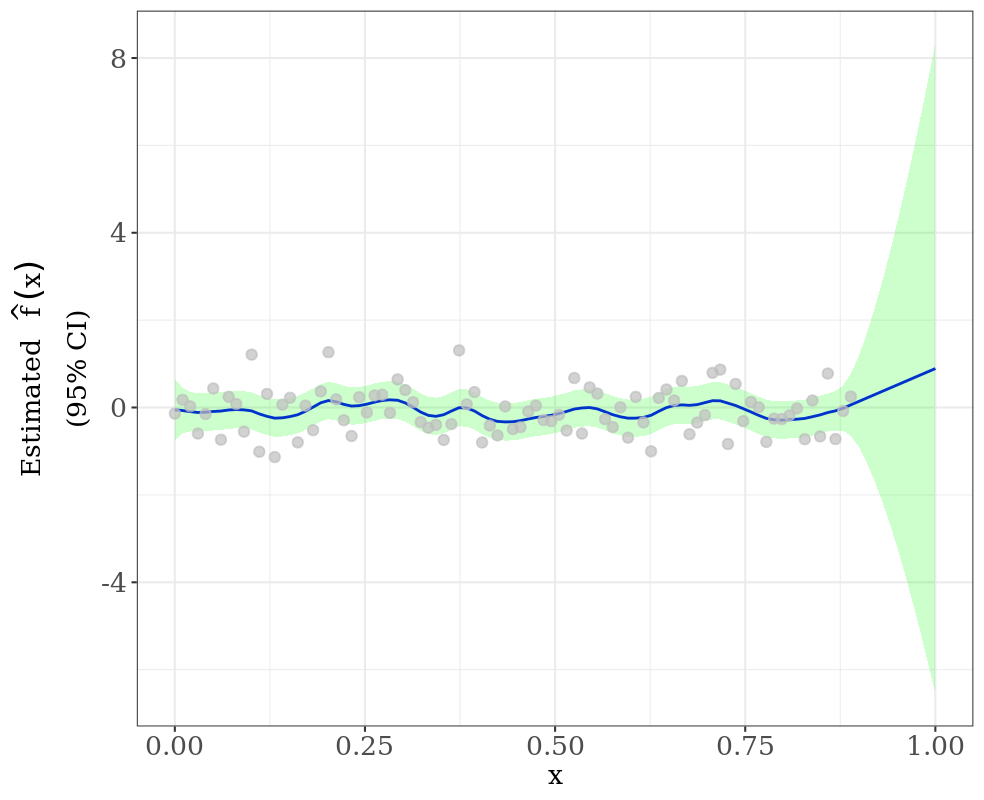}
    \caption{90\% observed locations}
    \label{fig:rw2-80}
  \end{subfigure}
  \hfill
  \begin{subfigure}[t]{0.32\textwidth}
    \centering
    \includegraphics[width=\linewidth]{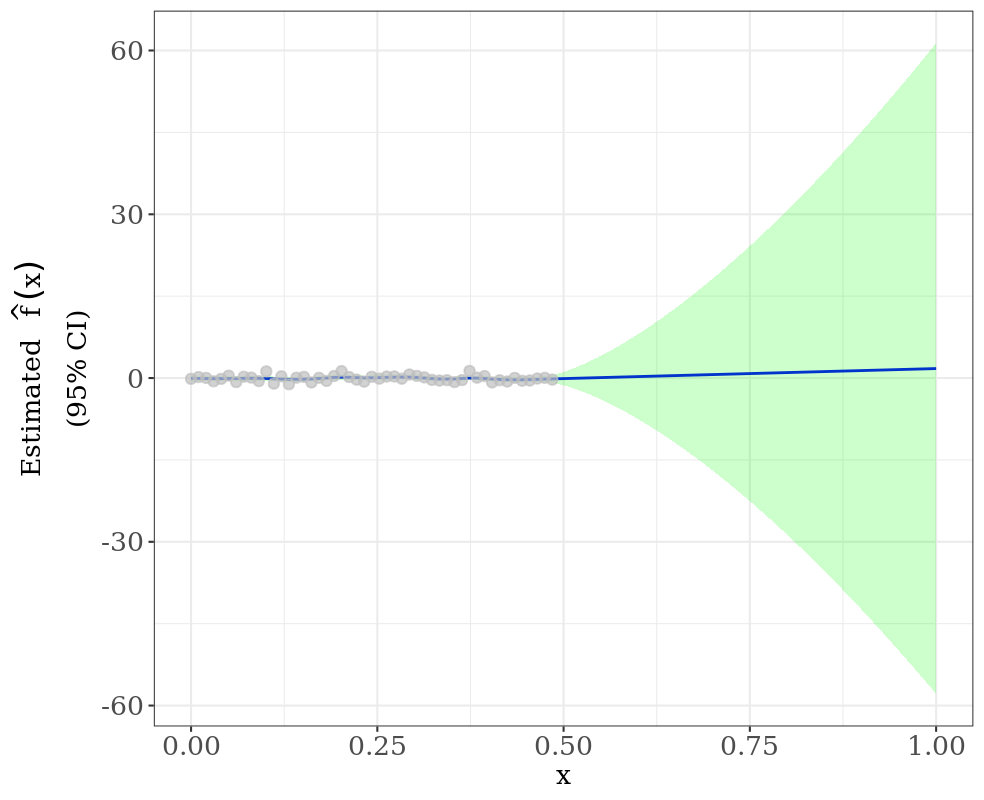}
    \caption{50\% observed locations}
    \label{fig:rw2-50}
  \end{subfigure}
  \hfill
  \begin{subfigure}[t]{0.32\textwidth}
    \centering
    \includegraphics[width=\linewidth]{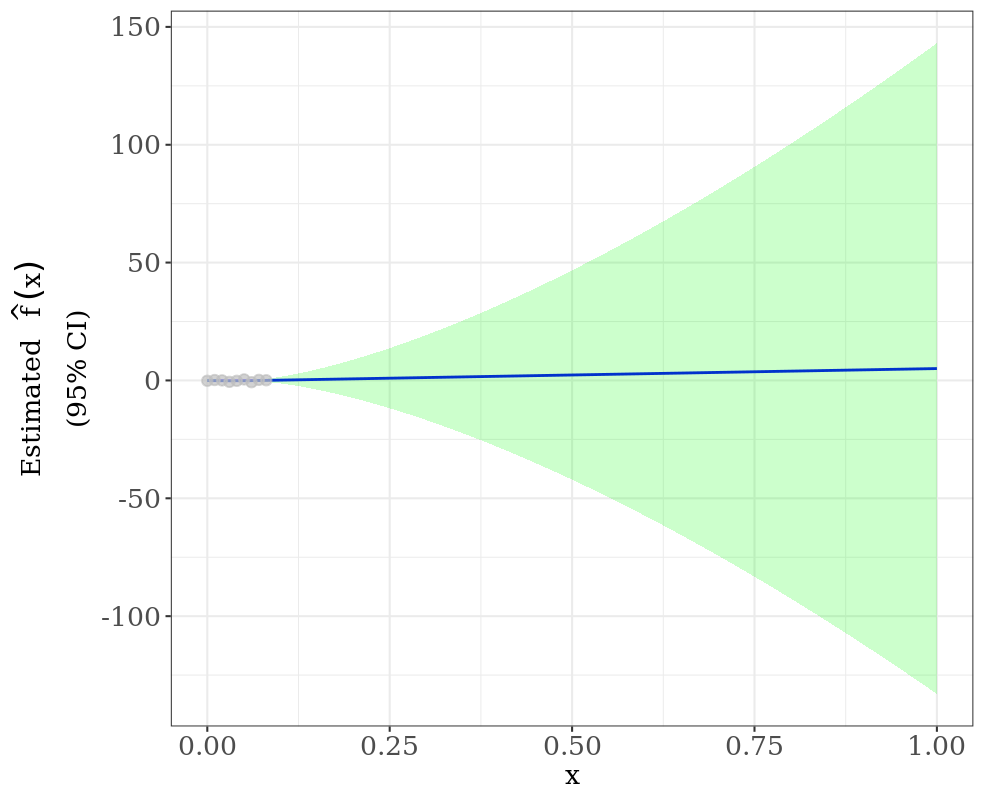}
    \caption{10\% observed locations }
    \label{fig:rw2-10}
  \end{subfigure}
\caption{Prediction and uncertainty of the RW2 model across three unobserved data scenarios.} 
  \label{fig:rw2-prediction-comparison}
\end{figure}

\paragraph{Predictive uncertainty grows without bound.}
In extrapolation beyond the observed range, intrinsic models have no stationary variance anchor outside the data. As a consequence, predictive uncertainty increases with the prediction horizon; for the RW2 model, the growth is cubic in distance, so credible intervals can expand extremely rapidly. Figure~\ref{fig:rw2-prediction-comparison} shows this behavior under increasing missingness: past the observed locations, the credible intervals widen until they far exceed the variation in the underlying signal.

\begin{figure}[ht]
  \centering
  \begin{subfigure}[b]{0.35\textwidth}
    \includegraphics[width=\textwidth]{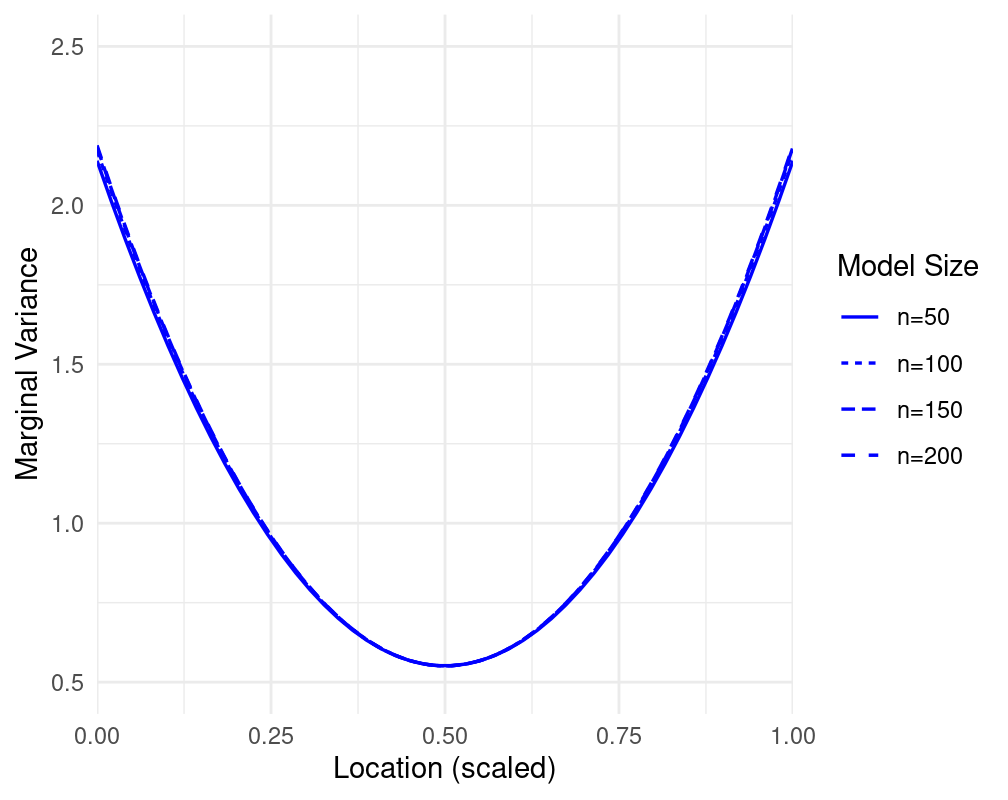}
    \caption{RW1}
    \label{fig:rw1-sensitivity}
  \end{subfigure}
  \hspace{0.02\textwidth}
  \begin{subfigure}[b]{0.35\textwidth}
    \includegraphics[width=\textwidth]{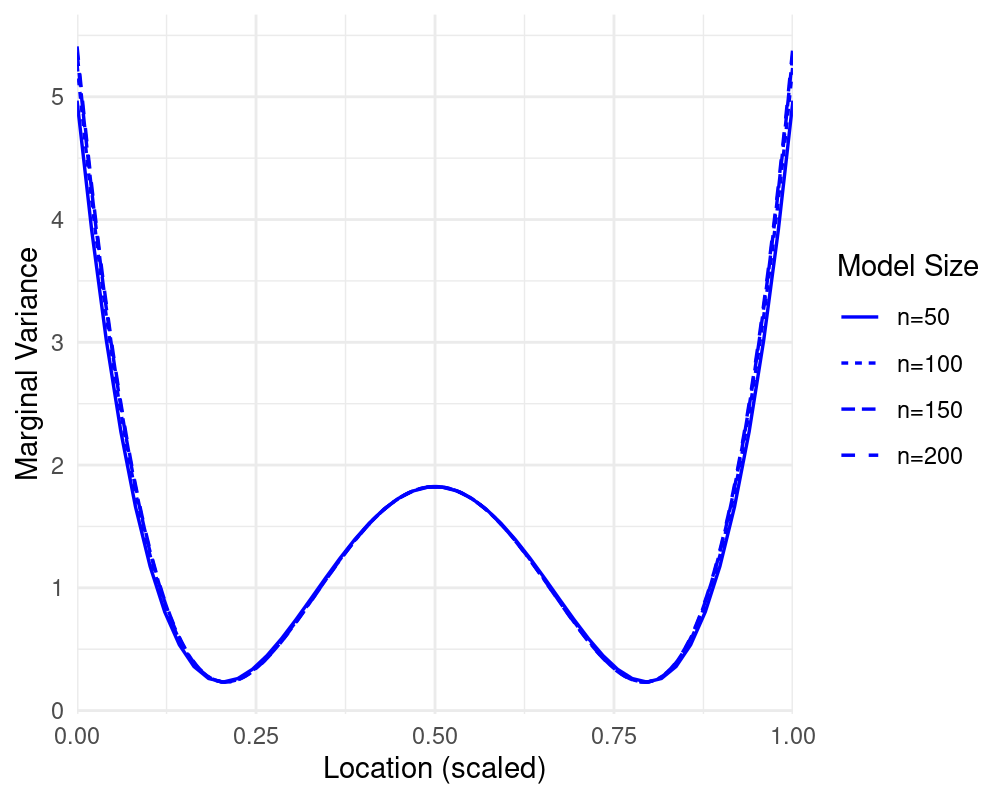}
    \caption{RW2}
    \label{fig:rw2-sensitivity}
  \end{subfigure}
  \caption{Scale-invariant marginal-variance distortion in the RW1 and RW2 models.}
  \label{fig:variance-sensitivity}
\end{figure}

\paragraph{Marginal variance is structurally non-constant, and the distortion is neither local nor removed by enlarging the domain.}
Intrinsicness also distorts uncertainty within the observed range: the implied marginal variance is not constant across locations, even when the model is intended to represent homogeneous smoothness. This heterogeneity is structural: it arises from intrinsicness rather than from local data support. In particular, random walk (RW) models typically exhibit inflated marginal variance near the boundaries, and this excess does not remain local: for higher-order models it propagates into the interior of the domain, affecting uncertainty well away from the endpoints. Moreover, this distortion is not a finite-domain artifact that disappears as the grid is enlarged. After suitable normalization, the marginal variance profile retains a stable shape as a function of relative location, so that enlarging the domain alters the scale but not the underlying pattern of variance distortion. Figure~\ref{fig:variance-sensitivity} displays this behavior for the RW1 and RW2 models across several domain lengths. The implication is that posterior uncertainty is shaped not only by the information in the observations but also by arbitrary choices in the computational representation of the domain --- the grid length, discretization, and boundary placement. A formal explanation of the marginal variance behavior based on explicit inverse formulas and the continuous Green's function limit is provided in Appendix~\ref{app:variance}.

\paragraph{Existing remedies do not remove impropriety.} These behaviors are recognized and several remedies have been proposed. The most common is variance rescaling, which standardizes the marginal variance of the intrinsic model so that the smoothing penalty has a consistent scale across domains (the \texttt{scale.model=TRUE} option in R-INLA). Such rescaling equalizes the marginal variances but addresses only the symptom: the underlying model remains intrinsic, so it is still rank-deficient, still requires global identifiability constraints, and still inherits the sensitivity of uncertainty to domain configuration and prediction horizon. The difficulty is therefore structural rather than one of scaling, and removing it requires replacing intrinsicness itself with a proper, stationary construction.

\subsection{A proper alternative} 

We propose the \emph{Proper Random Walk (PRW)} family. Its second-order member, the PRW2 model, is a finite-range GMRF with full-rank precision, constant marginal variance, and \emph{bounded} uncertainty for any admissible parameter values, while maintaining a sparse precision matrix for efficient inference. Here, \emph{proper} means a full-rank stationary model with an explicit range parameter, and should not be confused with variance-rescaled intrinsic RW2, which standardizes scale but remains rank-deficient. The PRW2 range parameter directly controls correlation decay; in the large-range regime, the PRW2 model closely reproduces RW2 smoothing, but remains proper throughout. We also provide a boundary correction for bounded domains and principled priors for the range and precision parameters.

\subsection{Outline of the paper} 

The remainder of the paper develops the PRW2 model and, in turn, demonstrates these properties. Section~\ref{sec:prw2-section} introduces the Proper Random Walk family, beginning with the first-order variant (PRW1) before developing PRW2 in detail: its discrete definition, continuous-time interpretation, GMRF construction via the Galerkin approximation, boundary correction, and prior specification. Section~\ref{sec:sim-study} describes a simulation study comparing the PRW2 and RW2 models in terms of predictive performance and uncertainty calibration, with particular attention to boundary prediction. Section~\ref{sec:app} provides a real-world application illustrating how PRW2's properties stabilize prediction in practice. In Section~\ref{sec:disc} we provide a discussion and connections to intrinsic formulations, including links to the Besag (ICAR) model and the broader role of proper stationary alternatives in Bayesian smoothing.

\section{Proper Random Walks} \label{sec:prw2-section}

We construct the PRW family as stationary full-rank generalizations of intrinsic RW models, with the aim of transforming intrinsically defined spline models into proper spline models that preserve local Markov structure while ensuring constant variance, coherent uncertainty quantification, and validity under both regular and irregular spacings.

\subsection{Proper Random Walk of Order 1 (PRW1)}

The \emph{Proper Random Walk of Order 1} (PRW1) is the first-order member of the PRW family. In continuous form, it is given by the Ornstein--Uhlenbeck (OU) process \citep{uhlenbeck1930theory}; under regular spacing, it reduces to the discrete AR(1) model \citep{jones1993longitudinal}. The two formulations are regarded as the continuous and discrete realizations of PRW1. This first-order construction is classical; we include it to establish the operator-based principle on which PRW2 is built.

Let \(\mathbf{x}=(x_1,\dots,x_T)^\top\) denote a Gaussian random vector, and let \(B\) denote the backshift operator, \(Bx_t=x_{t-1}\). The discrete-time PRW1 model is defined by
\[
(1-\rho B)x_t=\epsilon_t,
\qquad
\epsilon_t \sim N(0,\sigma_\epsilon^2),
\]
or equivalently, for \(t=2,\dots,T\),
\begin{equation}
    x_t=\rho x_{t-1}+\epsilon_t,
\qquad 0<\rho<1.
\label{ar1-equ}
\end{equation}

This is the classical first-order autoregressive model, AR(1). Its covariance function is \(r(h)=\operatorname{Cov}(x_t,x_{t-h})=\rho^{|h|}r(0),\) with marginal variance \(r(0)=\operatorname{Var}(x_t)=\frac{\sigma_\epsilon^2}{1-\rho^2}\). Thus, the PRW1 model has constant marginal variance and exponentially decaying dependence. For irregularly spaced locations, the natural continuous formulation is the zero-mean Ornstein--Uhlenbeck process. Formally, it may be written as
\[
x'(t)+\kappa x(t)=\sigma_w\,\eta(t),
\qquad \kappa>0,
\]
where \(\eta(t)\) denotes Gaussian white noise in the generalized sense and \(\sigma_w\) specifies its variance scale. Equivalently, \(\eta(t)\) can be viewed as the derivative of a standard Wiener process \(W(t)\), so that \( \eta(t)\,dt=dW(t) \), and the model can be written rigorously in differential form as
\(
dx(t)+\kappa x(t)\,dt=\sigma_w\,dW(t)
\). Its covariance function is
\(
\operatorname{Cov}\{x(s),x(t)\}
=
\frac{\sigma_w^2}{2\kappa} \exp \; \!\bigl(-\kappa|s-t|\bigr),
\)
so that
\(
\operatorname{Var}\{x(t)\}=\frac{\sigma_w^2}{2\kappa}
\). 

Moreover, for \(t_j>t_i\), we have
\(
x(t_j)=e^{-\kappa(t_j-t_i)}\;x(t_i) \;+ \;\eta_{ij}
\), where
\(
\eta_{ij}\sim N\!\left(
0,\,
\frac{\sigma_w^2}{2\kappa}\left[1-e^{-2\kappa(t_j-t_i)}\right]
\right)
\). When the process is sampled on a regular grid with spacing \(\delta =1\), we get the AR(1) process given in \eqref{ar1-equ} with \(
\rho=e^{-\kappa\delta},
\) and
\(
\epsilon_t \sim N\!\left(
0,\,
\frac{\sigma_w^2}{2\kappa}(1-\rho^2)
\right)
\). These are the standard continuous-to-discrete relations for the OU and AR(1) models \citep{jones1993longitudinal}. The continuous formulation is therefore the natural Gaussian, Markov, and proper analog for irregularly spaced locations.

The PRW1 model already illustrates the main technical advantages of proper construction: constant marginal variance. Moreover, the prediction variance over an interval \((t_i,t_j)\) is \(\operatorname{Var}\{x(t_j)\mid x(t_i)\}=\frac{\sigma_w^2}{2\kappa}\left(1-e^{-2\kappa(t_j-t_i)}\right),\) which is bounded above by the marginal variance \(\sigma_w^2/(2\kappa)\). Thus, even when the uncertainty increases with separation from the last observation, it remains bounded. 
For comparison, the intrinsic RW1 model is defined by
\[
x_t=x_{t-1}+\epsilon_t,
\qquad
\epsilon_t\sim N(0,\sigma_\epsilon^2),
\]
so that an \(h\)-step-ahead prediction satisfies
\(
x_{t+h}=x_t+\sum_{k=1}^{h}\epsilon_{t+k}
\). Consequently, \(\operatorname{Var}(x_{t+h}\mid x_t)=h\,\sigma_\epsilon^2,\) increases linearly with the prediction distance and has no finite upper bound. Thus, unlike the PRW1 model, the intrinsic RW1 model does not merely exhibit increasing uncertainty away from the last observation; it accumulates uncertainty without bound. Next we extend the construction from the first order to the second order. This second-order extension follows the same operator logic that links RW1 to RW2. Rather than appealing to the general second-order autoregressive class, AR(2), $x_t=\phi_1x_{t-1}+\phi_2x_{t-2}+\varepsilon_t$, we construct the proper second-order spline model by iterating the proper first-order operator. This preserves the structural analogy with RW2 and leads directly to PRW2.

\subsection{The Second-Order Proper Random Walk (PRW2)}\label{sec:prw2}

To make the PRW2 construction precise, it is helpful to first see explicitly 
how RW2 arises from RW1 by operator composition. RW1 is defined by the 
first-order difference operator
\[
(1-B)\,x_t = \epsilon_t, \qquad \epsilon_t \sim \mathcal{N}(0,\sigma_\epsilon^2),
\]
and RW2 is obtained by applying this same operator a second time to the 
driving noise: if $\epsilon_t$ itself follows a RW1, i.e.\
$(1-B)\,\epsilon_t = \varepsilon_t$, so that $\epsilon_t = (1-B)^{-1}\varepsilon_t$, 
then substituting gives
\[
(1-B)\,x_t = (1-B)^{-1}\varepsilon_t \;\implies\; (1-B)^2\,x_t = \varepsilon_t,
\]
which is precisely the RW2 model. PRW2 is obtained by the same construction, 
replacing the intrinsic operator $(1-B)$ with its proper counterpart 
$(1-\rho B)$ from \eqref{ar1-equ}. If $\epsilon_t$ itself follows a 
PRW1, i.e.\ $(1-\rho B)\,\epsilon_t = \varepsilon_t$, so that 
$\epsilon_t = (1-\rho B)^{-1}\varepsilon_t$, then substituting produces
\[
(1-\rho B)\,x_t = (1-\rho B)^{-1}\varepsilon_t 
\;\implies\; 
(1-\rho B)^2\,x_t = \varepsilon_t.
\]
Formally,
\begin{equation}
(1-\rho B)^2 x_t = \varepsilon_t,
\qquad \varepsilon_t \sim \mathcal{N}(0,\sigma_\varepsilon^2),
\qquad 0 < \rho < 1,
\label{eq:prw2-operator}
\end{equation}
which makes the analogy with RW2 completely transparent: PRW2 is to PRW1 
exactly what RW2 is to RW1---a second-order model obtained by composing two 
first-order operators, with the only structural difference being the 
replacement of $(1-B)$ with the proper stationary operator $(1-\rho B)$.
Expanding \eqref{eq:prw2-operator} gives the explicit recursion
\begin{equation}
x_t = 2\rho\, x_{t-1} - \rho^2\, x_{t-2} + \varepsilon_t,
\label{eq:prw2-disc}
\end{equation}
which has the algebraic form of an AR(2) process with coefficients
$\phi_1 = 2\rho$ and $\phi_2 = -\rho^2$. Both coefficients are governed by
the single parameter $\rho$, so PRW2 is not the general two-parameter AR(2)
model but the degenerate, coincident-root case that our operator construction
$(1-\rho B)^2$ singles out. This restriction is what gives the model its
spline-like smoothing structure, and it defines a non-intrinsic (proper)
second-order Gaussian Markov random field, viewed here as a PRW spline model.
As in PRW1, $\rho$ controls the dependence scale while $\sigma_\varepsilon$
determines the innovation variance. The covariance function is
\[
r(h) = \sigma^{2}\!\left(1+h\frac{1-\rho^{2}}{1+\rho^{2}}\right)\rho^{h}, \qquad h \ge 0,
\] where $\sigma^{2}=r(0)=\sigma_\varepsilon^{2}\,\dfrac{1+\rho^{2}}{(1-\rho^{2})^{3}}$ is the marginal variance.
  
The precision matrix $\mathbf Q$ of a random vector $\mathbf x=(x_1,\dots,x_n)^\top$ following a PRW2 process is pentadiagonal and nonsingular, with determinant $\det(\mathbf Q)=(1-\rho^2)^4\sigma_\varepsilon^{-2n}$. The banded structure reflects the Markov property of the model: conditionally on the past, each value depends only on its two nearest predecessors.

\subsubsection{Continuous-Time Representation}

Following the standard connection between autoregressive difference equations and their continuous time analogs, we represent the PRW2 in continuous time using the second-order stochastic differential equation (SDE) corresponding to its discrete operator form \citep{priestley1981spectral, jones1993longitudinal}. Using the same generalized white-noise notation introduced for PRW1, continuous-time PRW2 satisfies
\begin{equation}
x''(t) + 2\kappa x'(t) + \kappa^{2}x(t) = \sigma_w\,\eta(t),
\label{eq:prw2-sde}
\end{equation}
Sampling the continuous process on a regular grid of spacing $\delta$ reproduces the discrete model in \eqref{eq:prw2-disc} to first order in $\kappa\delta$, with the two coinciding as $\delta\to 0$. The stationary covariance is
\[
r(h) = \frac{\sigma_w^{2}}{4\kappa^{3}}\,(1+\kappa h)\,e^{-\kappa h}, \qquad h \ge 0,
\]
with marginal variance $r(0)=\sigma_w^2/(4\kappa^3)$. The key parameter is \(\kappa\), which is most naturally interpreted through the covariance structure: small \(\kappa\) produces slowly decaying correlations, whereas large \(\kappa\) produces shorter-range dependence. A more interpretable parameterization of the model is via the correlation range \(\varrho\), defined as the distance at which the correlation becomes practically negligible. In the present setting,
\(
\varrho=\frac{2\sqrt{3}}{\kappa},
\)
obtained from the general Mat\'ern parameterization
\(
\varrho=\frac{\sqrt{8\nu}}{\kappa},
\)
with \(\nu=3/2\), where \(\nu\) denotes the Mat\'ern smoothness parameter. This value, \(\nu=3/2\), is the smoothness associated with the second-order model; by comparison, first-order models such as RW1 and PRW1 correspond to \(\nu=1/2\). In the parameterization of \citet{lindgren2011explicit}, also used by \citet{fuglstad2019constructing}, \(\varrho\) is approximately the distance at which the correlation drops to \(0.1\). 

Three parameterizations therefore describe the model---exactly in continuous
time and to first order in $\kappa\delta$ on a regular grid---and it is
convenient to collect their correspondence in one place. The discrete
recursion is governed by the dependence parameter $\rho$, the continuous SDE by
the rate $\kappa$, and the covariance by the range $\varrho$. On a regular grid
of spacing $\delta$, the discrete dependence parameter inherits the same
continuous--discrete relation as in PRW1, $\rho = e^{-\kappa\delta}$; combining
this with the range relation $\varrho = 2\sqrt{3}/\kappa$ collapses the three
descriptions into a single chain,
\begin{equation}
\kappa = \frac{2\sqrt{3}}{\varrho},
\qquad
\rho = e^{-\kappa\delta} = \exp\!\left(-\frac{2\sqrt{3}\,\delta}{\varrho}\right).
\label{eq:param-link}
\end{equation}
Each map in \eqref{eq:param-link} is strictly monotone, so the three parameters
carry the same information and may be used interchangeably; in particular,
\begin{equation}
\rho \to 1
\quad\Longleftrightarrow\quad
\kappa \to 0
\quad\Longleftrightarrow\quad
\varrho \to \infty,
\label{eq:param-limit}
\end{equation}
which is the limit examined in Section~\ref{sec:rw2-limit}. In practice we use
$\rho$ in discrete derivations, $\kappa$ in the SDE and the finite-element
construction, and $\varrho$ for interpretation and prior specification, since
the SDE is written most naturally in terms of $\kappa$ while $\varrho$ provides
the more interpretable scale.

The continuous-time formulation is also important beyond interpretation. It provides the basis for numerical approximations on irregular grids, allowing the model to be extended beyond regularly spaced observations, while retaining the same local Markov structure as in the discrete formulation.

\subsubsection{RW2 as a limiting case} \label{sec:rw2-limit}

A key design feature of the PRW2 model is that it retains a direct connection
to the classical RW2 spline model. As seen from the defining operator
\eqref{eq:prw2-operator}, letting \(\rho \to 1\) recovers the second-order
intrinsic differencing operator \((1-B)^2\), so that the RW2 structure arises
as the limiting case of the proper model \((1-\rho B)^2\).

The same conclusion is immediate in the continuous parameterization. By the
maps in \eqref{eq:param-link}, the single limit \eqref{eq:param-limit} sends
$\rho \to 1$, $\kappa \to 0$, and $\varrho \to \infty$ together, so the RW2
model appears as the infinite-range limit of the PRW2 model. In this limit, the
restoring effect of the proper model vanishes, recovering the intrinsic
second-order model.

This limiting relationship is fundamental for hyperparameter prior specification. Because the RW2 model sits at the boundary of the family, it is the natural base model: the prior preserves RW2-type smoothing behavior by default, while allowing the data to pull toward a proper finite-range model when warranted.

\subsubsection{Penalized Complexity Priors}\label{pc}

We derive penalized complexity (PC) priors \citep{simpson2017penalising} for
the range $\varrho$ and precision $\tau$, assuming prior
independence $\pi(\varrho,\tau)=\pi(\varrho)\,\pi(\tau)$. In the PC
framework, the prior is constructed by placing an exponential distribution on
the Kullback--Leibler distance $d(\theta)=\sqrt{2\,\mathrm{KLD}(p_\theta\|p_{\theta_0})}$
between the model at $\theta$ and a simpler base model at $\theta_0$, giving
$\pi(\theta)\propto\exp(-\lambda\,d(\theta))\,|\partial d/\partial\theta|$.
The rate $\lambda>0$ is calibrated through a tail-probability statement on
the parameter scale.

\paragraph{PC prior for the range.}

The natural base model for the PRW2 model is the RW2 limit when $\rho=1$. Taking $\rho_0=1$ as the base formulation, the KL distance
between the PRW2 model and the RW2 limit reduces to
\[
d(\rho) = \sqrt{3 - 4\gamma_1 + \gamma_2}, \qquad 0 < \rho < 1,
\]
where $\gamma_1=2\rho/(1+\rho^2)$ and $\gamma_2=(3\rho^2-\rho^4)/(1+\rho^2)$
are the lag-1 and lag-2 correlations of the PRW2 process. Since $d(\rho)$
ranges in $(0,\sqrt{3})$ as 
$\rho$ ranges in $(0,1)$, the exponential prior on $d$ is truncated,
and the PC prior on $\rho$ is
\[
\pi(\rho) =
\frac{\lambda\,\exp\!\bigl(-\lambda\sqrt{3-4\gamma_1+\gamma_2}\bigr)}
     {1-\exp(-\sqrt{3}\,\lambda)}
\cdot
\frac{\rho^5+2\rho^3-4\rho^2-3\rho+4}
     {(1+\rho^2)^2\sqrt{3-4\gamma_1+\gamma_2}},
\qquad 0<\rho<1.
\]
To express the prior on the interpretable range scale, we use the continuous--discrete link \eqref{eq:param-link}, $\rho = \exp(-2\sqrt{3}\,\delta/\varrho)$, so that $\varrho\to\infty$ corresponds to $\rho\to 1$ (the RW2 base model). The Jacobian of this 
reparameterization is
\[
\left|\frac{\partial\rho}{\partial\varrho}\right|
= \frac{2\sqrt{3}\,\delta}{\varrho^2}
  \exp\!\left(-\frac{2\sqrt{3}\,\delta}{\varrho}\right),
\]
and the prior on $\varrho$ follows by the standard change of variables
$\pi(\varrho)=\pi(\rho(\varrho))\,|\partial\rho/\partial\varrho|$.
Any multiplicative constant in $d$, such as the factor depending on $n$ that arises from accumulating the divergence over the grid, is absorbed into the calibration of the rate parameter $\lambda$ rather than appearing explicitly in the prior on $\varrho$. Since $d$ depends on $\varrho$ only through $\delta/\varrho$, the induced prior is approximately invariant to $n$ for fine grids ($\varrho \gg \delta$). The rate $\lambda$ is calibrated through an interpretable tail probability on the range scale,
\[
\Pr(\varrho < \varrho_0) = \alpha_\varrho,
\]
with the threshold $\varrho_0$ taken as a fraction of the domain length. This
lets the user control how much prior mass favors long range, smooth behavior; the calibration used in our experiments is described in Section~\ref{sec:sim-design}. Because the threshold scales with the domain, the same calibration applies, to leading order, regardless of the number of observations or the grid resolution.     

The tail behaviour of the prior on $\varrho$ is critical: the following result
shows that priors whose tails decay too quickly (so that
$\mathbb{E}[\varrho^{3/2}]<\infty$) cannot shrink towards the RW2 base model.

  \begin{theorem}
Let $\pi_\varrho(\varrho)$ be a proper prior on $\varrho$ whose density is
eventually non-increasing. Since
$d(\rho)^2=(1-\rho)^3(\rho+3)/(1+\rho^2)$ and
$1-\rho=2\sqrt{3}\,\delta/\varrho+O(\varrho^{-2})$, we have
$d(\varrho)=c\,\varrho^{-3/2}+O(\varrho^{-5/2})$ as $\varrho\to\infty$, with $c=\sqrt{2}\,(2\sqrt{3}\,\delta)^{3/2}$. Consequently, if
$\mathbb{E}[\varrho^{3/2}]<\infty$, then the induced prior on the PC distance
satisfies $\pi_d(d)\to 0$ as $d\to 0$.
\end{theorem}

\noindent This means any prior on $\varrho$ with a finite $3/2$-moment assigns
zero density at the RW2 base model, preventing the posterior from concentrating
on the correct model even when it is the true one. The PC prior avoids this by
construction: its density is maximized at $d=0$, penalizing deviations from the
RW2 model at a constant rate. Its tail, $\pi(\varrho)\propto\varrho^{-5/2}$,
gives $\mathbb{E}[\varrho^{3/2}]=\infty$, so it lies exactly at the boundary of
the theorem. The rate $d\propto\varrho^{-3/2}$ coincides with that of a Mat\'ern
field in three dimensions, $d\propto\varrho^{-D/2}$ \citep{fuglstad2019constructing};
for PRW2 it arises from the $(1-\rho)^3$ factorization rather than from the
spatial dimension.


\begin{figure}[H]
\centering

\includegraphics[width=0.95\linewidth]{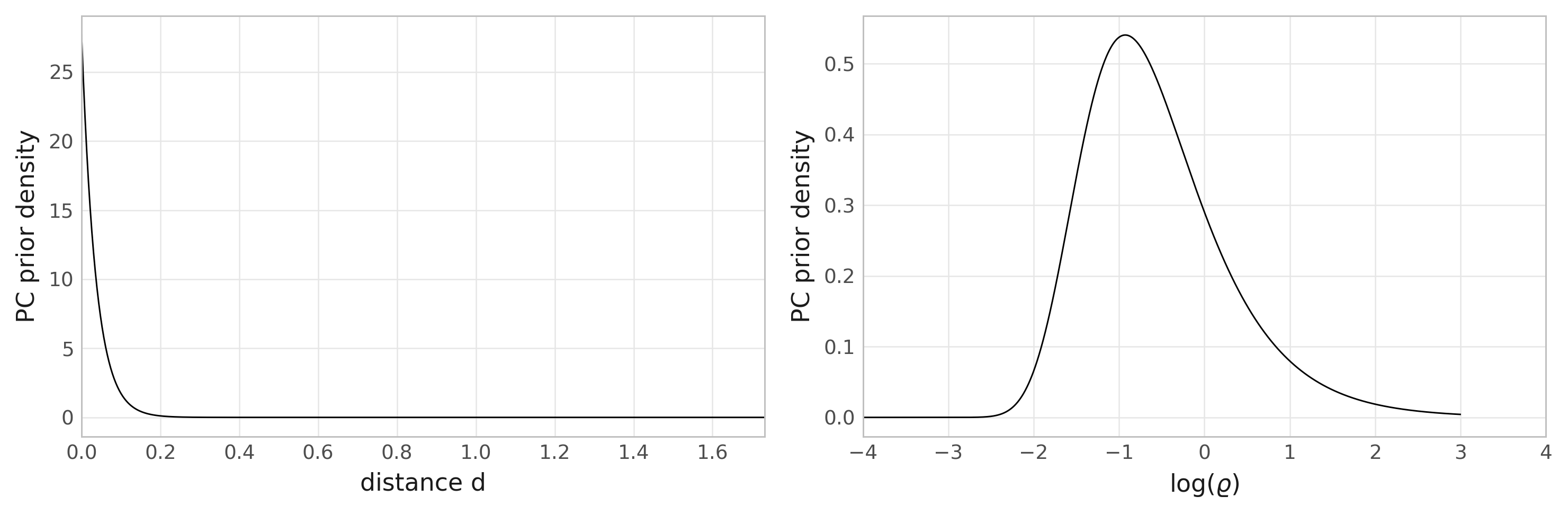}
\caption{Penalized complexity prior for the PRW2 range. On the distance scale (left) the prior is a truncated exponential that concentrates at $d=0$, the RW2 base model; the prior on the log-range scale (right) is shown for the corresponding parametrization used in INLA.}
\label{fig:pc-prior}
\end{figure}

\paragraph{PC prior for precision.}

For the precision parameter $\tau=1/\sigma^2$, the base model is ``no effect''
($\sigma=0$, $\tau\to\infty$). Following \citet{simpson2017penalising}, an
exponential prior for the marginal standard deviation $\sigma=\tau^{-1/2}$
induces
\[
\pi(\tau)=\frac{\lambda_\tau}{2}\,\tau^{-3/2}
\exp\!\bigl(-\lambda_\tau\,\tau^{-1/2}\bigr), \qquad \tau>0,
\]
calibrated via $\Pr(\sigma>u)=\alpha_\sigma$, giving
$\lambda_\tau=-\log(\alpha_\sigma)/u$.

\subsubsection{GMRF Construction via the Galerkin Approximation}\label{sec:galerkin}
The continuous-time representation \eqref{eq:prw2-sde} can be projected onto a finite-dimensional subspace using the finite-element method (FEM) through its Galerkin formulation. This approach provides a direct bridge between the stochastic differential equation and its GMRF representation \citep{lindgren2011explicit}. Let $\{\psi_i(t)\}_{i=1}^n$ denote a set of piecewise linear (“hat”) basis functions that span the approximation space, and express the process as $x(t) \approx \sum_{i=1}^n x_i \psi_i(t)$, where the coefficients $\{x_i\}$ define the latent random field.
Substituting this expansion into the SDE and multiplying by each test function $\psi_i$, the weak form of the equation is
\[
\langle \psi_i, L(x) \rangle = \langle \psi_i, \sigma_w\,\mathrm{d}B(t) \rangle, \qquad i=1,\dots,n,
\]
where the inner product is defined as $\langle f,g\rangle = \int f(t)g(t)\,\mathrm{d}t$ and $L(x)=x''(t)+2\kappa x'(t)+\kappa^2 x(t)$. The Galerkin discretization leads to the linear system $\mathbf{Kx}=\mathbf{W}$, where $\mathbf{W}$ is a Gaussian vector induced by the stochastic forcing term and $\mathbf{K}$ is the global system matrix.

The system matrix can be expressed as
\[
\mathbf{K} = \kappa^{2}\mathbf{C} + 2\kappa\,\mathbf{M} - \mathbf{G},
\] where the component matrices are
\[
C_{ij} = \langle \psi_i, \psi_j \rangle, \qquad 
M_{ij} = \langle \psi_i, \psi_j' \rangle, \qquad
G_{ij} = \langle \psi_i', \psi_j' \rangle.
\]
Here $\mathbf{C}$ is the (tridiagonal) mass matrix, $\mathbf{M}$ is the first-derivative (convection) coupling, and $\mathbf{G}$ is the stiffness matrix, which --- via integration by parts under the homogeneous boundary conditions, $\langle \psi_i, \psi_j''\rangle = -\langle \psi_i', \psi_j'\rangle$ --- encodes the second-derivative (roughness) term. Together these assemble the weak form of the operator $L(x)=x''(t)+2\kappa x'(t)+\kappa^2 x(t)$.

The random coefficients $\mathbf{x}=(x_1,\dots,x_n)^\top$ follow a Gaussian distribution with precision matrix
\[
\mathbf{Q} = \frac{\tau}{4\kappa^3}\,\mathbf{K}^\top\tilde{\mathbf{C}}^{-1}\mathbf{K},
\]
where $\tau$ is the marginal precision of the field (the factor $1/(4\kappa^3)$
normalizes the stationary variance $\sigma_w^2/(4\kappa^3)$ to $1/\tau$) and
$\tilde{\mathbf{C}}$ is the lumped (diagonal) mass matrix,
$\tilde C_{ii}=\sum_j C_{ij}$; on a regular grid of spacing $\delta$,
$\tilde{\mathbf{C}}=\delta\,\mathbf{I}$ apart from the two end nodes. Mass
lumping replaces the dense inverse of $\mathbf{C}$ by a diagonal one. The
induced modification to the stationary covariance is negligible, while the
resulting precision matrix remains sparse and computationally efficient
\citep{lindgren2011explicit}.

The Gaussian vector $\mathbf{x}$ therefore represents a finite-dimensional approximation of the continuous PRW2 process. Its precision matrix $\mathbf{Q}$ encodes the same local dependence implied by the discrete recursion in \eqref{eq:prw2-disc}: each coefficient $x_i$ interacts only with its nearest neighbors through the overlapping basis functions. Consequently, the Galerkin construction reproduces the Markov property of the PRW2 model in a continuous setting, providing a unified formulation in which the discrete and continuous representations coincide as the grid spacing $\delta \to 0$.

\subsubsection{Stationarity and Boundary Corrections}\label{sec:boundary}

\begin{figure}[h!]
\centering
\begin{minipage}[b]{0.48\textwidth}
    \centering
    \includegraphics[width=\linewidth,height=7cm]{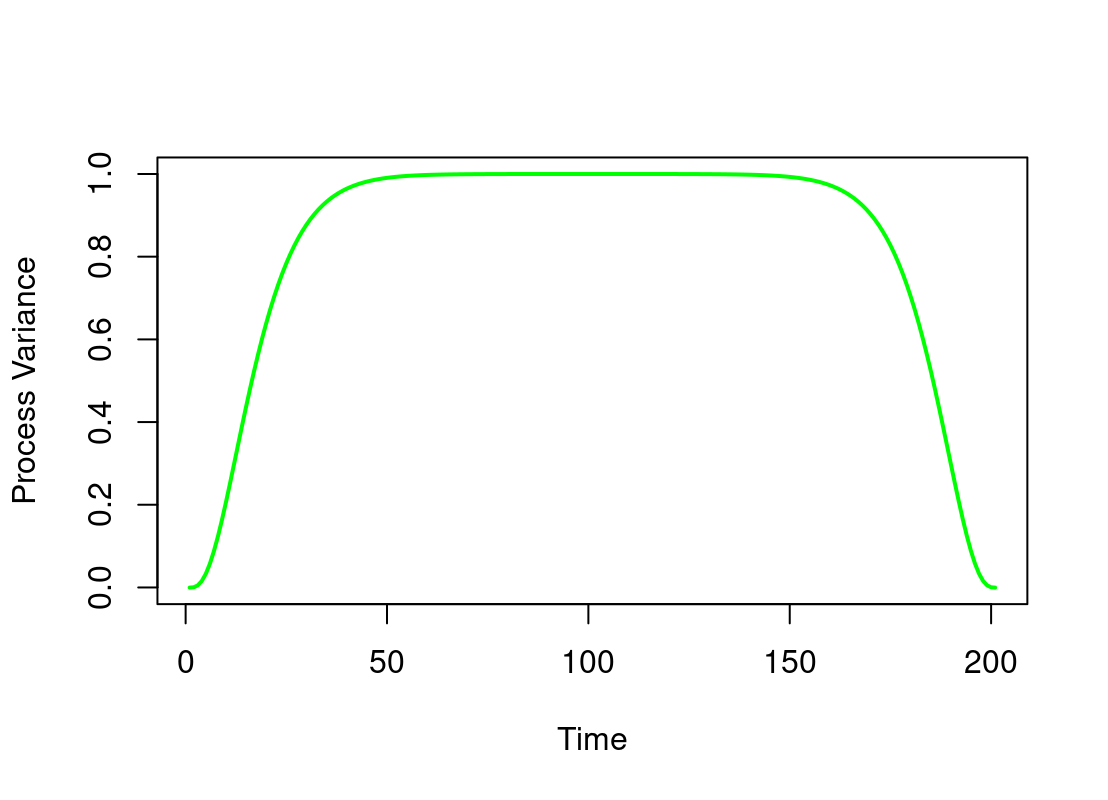}
    \caption*{(a) Without boundary correction}
\end{minipage}
\hfill
\begin{minipage}[b]{0.48\textwidth}
    \centering
    \includegraphics[width=\linewidth,height=7cm]{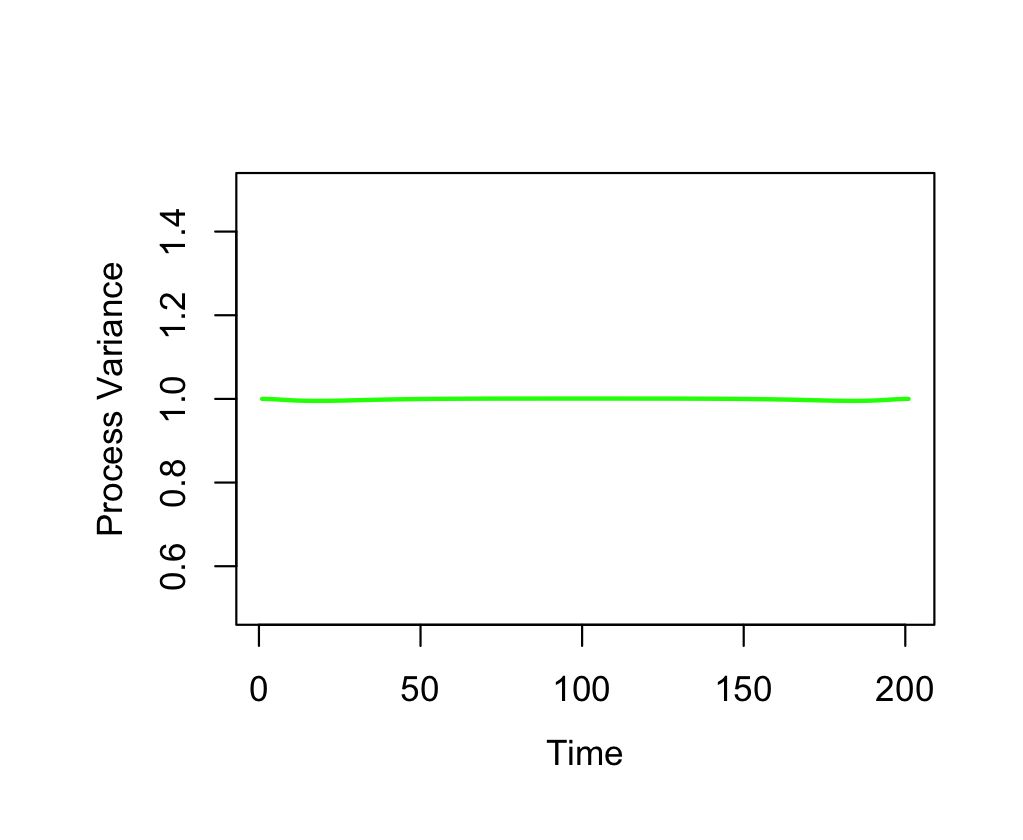}
    \caption*{(b) With boundary correction}
\end{minipage}
\caption{Boundary correction recovers approximately constant marginal variance across the domain.}
\label{fig:boundary_correction_comparison}
\end{figure}

In bounded domains, the Galerkin discretization of the SDE requires boundary conditions to obtain a well-posed system. Deterministic constraints (e.g., Dirichlet-type conditions) can distort the stationary covariance near the edges by breaking the stationarity implied by the underlying PRW2 process (Figure~\ref{fig:boundary_correction_comparison}a). To restore stationarity, we directly correct the boundary block of the Galerkin precision matrix so that the boundary nodes follow the stationary joint law of the process. Concretely, the correction is applied to the first and last two rows and columns of $\mathbf{Q}$.

Let $\mathbf{x}=(\mathbf{x}_\partial^\top,\mathbf{x}_B^\top)^\top$, where
\[
\mathbf{x}_\partial = (x_1, x_2, x_{n-1}, x_n)^\top, 
\qquad 
\mathbf{x}_B = (x_3, x_4, \ldots, x_{n-2})^\top.
\]
Write the corresponding block partition of the Galerkin precision matrix as
\[
\mathbf{Q}=
\begin{bmatrix}
\mathbf{Q}_{\partial\partial} & \mathbf{Q}_{\partial B}\\[3pt]
\mathbf{Q}_{B\partial} & \mathbf{Q}_{BB}
\end{bmatrix},
\]
and let $\boldsymbol{\Sigma}_{\partial\partial}$ denote the \emph{stationary} covariance of $\mathbf{x}_\partial$ implied by the PRW2 model, assembled from the closed-form stationary covariance $r(|\Delta|)=(\sigma_w^2/4\kappa^3)(1+\kappa|\Delta|)e^{-\kappa|\Delta|}$ evaluated at the lags between the four boundary nodes. Enforcing
$\pi(\mathbf{x}_\partial)\sim\mathcal{N}(0,\boldsymbol{\Sigma}_{\partial\partial})$ while retaining the same conditional model for the interior, $\pi(\mathbf{x}_B\mid \mathbf{x}_\partial)$, yields the corrected boundary precision block
\[
\mathbf{Q}^{\mathrm{corr}}_{\partial\partial}
= \boldsymbol{\Sigma}_{\partial\partial}^{-1} + \mathbf{Q}_{\partial B}\mathbf{Q}_{BB}^{-1}\mathbf{Q}_{B\partial}.
\]
This formulation corrects both boundaries simultaneously and preserves the cross-covariance between the left and right boundary nodes through $\boldsymbol{\Sigma}_{\partial\partial}$. The interior block $\mathbf{Q}_{BB}$ is banded, so the required solves use sparse Cholesky factorization; the correction adds only a low-rank coupling between the two boundary blocks---negligible when $\varrho$ is small relative to the domain length---so the corrected precision remains sparse and the overall factorization cost is $O(n)$. Figure~\ref{fig:boundary_correction_comparison} shows that, after correction, the discretized field recovers stationarity and approximately constant marginal variance across the bounded domain.

\section{Simulation study: PRW2 versus RW2 for smoothing and prediction}
\label{sec:sim-study}

We compare the PRW2 with the intrinsic RW2 spline models in controlled one-dimensional experiments, focusing on whether the PRW2 reproduces RW2 spline-like smoothing when data are informative, while providing stable, bounded uncertainty in sparse regions (gaps and domain boundaries). Both models are fitted with the same Gaussian observation model, with hyperparameter priors as described in Section~\ref{pc}.

\subsection{Design}
\label{sec:sim-design}
Data are generated from
\[
y_i = f(x_i) + \epsilon_i,\qquad \epsilon_i \sim \mathcal{N}(0,\sigma_y^2),
\]
on a regular grid of $n=100$ locations $x_i\in[0,1]$.
We consider two underlying signals:
\[
f_{\text{struct}}(x)=\{\sin(2.5\pi x)\}^3,
\qquad
f_{\text{flat}}(x)\equiv 0.
\]
To assess both the fitting and the prediction behavior within a unified framework, we study three data-availability regimes:
(i) a dense setting with all observations available;
(ii) mid-domain interpolation with a contiguous block of missing responses (the gap regime, where $y_{31:50}=\texttt{NA}$);
and (iii) prediction of the right-edge with the final portion of the domain treated as unobserved (the edge regime, where $y_{81:100}=\texttt{NA}$).
Predictions are obtained by treating withheld locations as missing responses, which yield posterior summaries for the latent field at those locations.
For the gap and edge regimes, the shaded region in Figures~\ref{fig:gap-pred} -- \ref{fig:edge-pred} denotes the prediction region used for the calculation of summary metrics.
Results are averaged over repeated simulations (50 replicates). Both the PRW2 and RW2 latent fields are assigned penalized complexity priors.
The marginal standard deviation $\sigma$ is given a PC prior calibrated by
$\Pr(\sigma>0.5)=0.01$ (the \texttt{R-INLA} setting \texttt{pc.prec} with
\texttt{param=c(0.5,0.01)}), applied identically to both models. The PRW2 range
is given a PC prior calibrated by $\Pr(\varrho<L/2)=0.1$, where $L$ is the domain
length, placing most prior mass on long-range (RW2-like) behaviour; here $L=1$,
so $\varrho_0=0.5$. The observation noise standard deviation is held fixed at $\sigma_y = 0.2$ in both models.

\subsection{Results}
\label{sec:sim-results}

We report (i) the root mean square error (RMSE) of the posterior mean and (ii) the mean width of the $95\%$ credible intervals (CIs). For the gap and edge regimes, metrics are computed only on the corresponding prediction region.

\begin{figure}[H]
\centering
\includegraphics[width=0.6\linewidth]{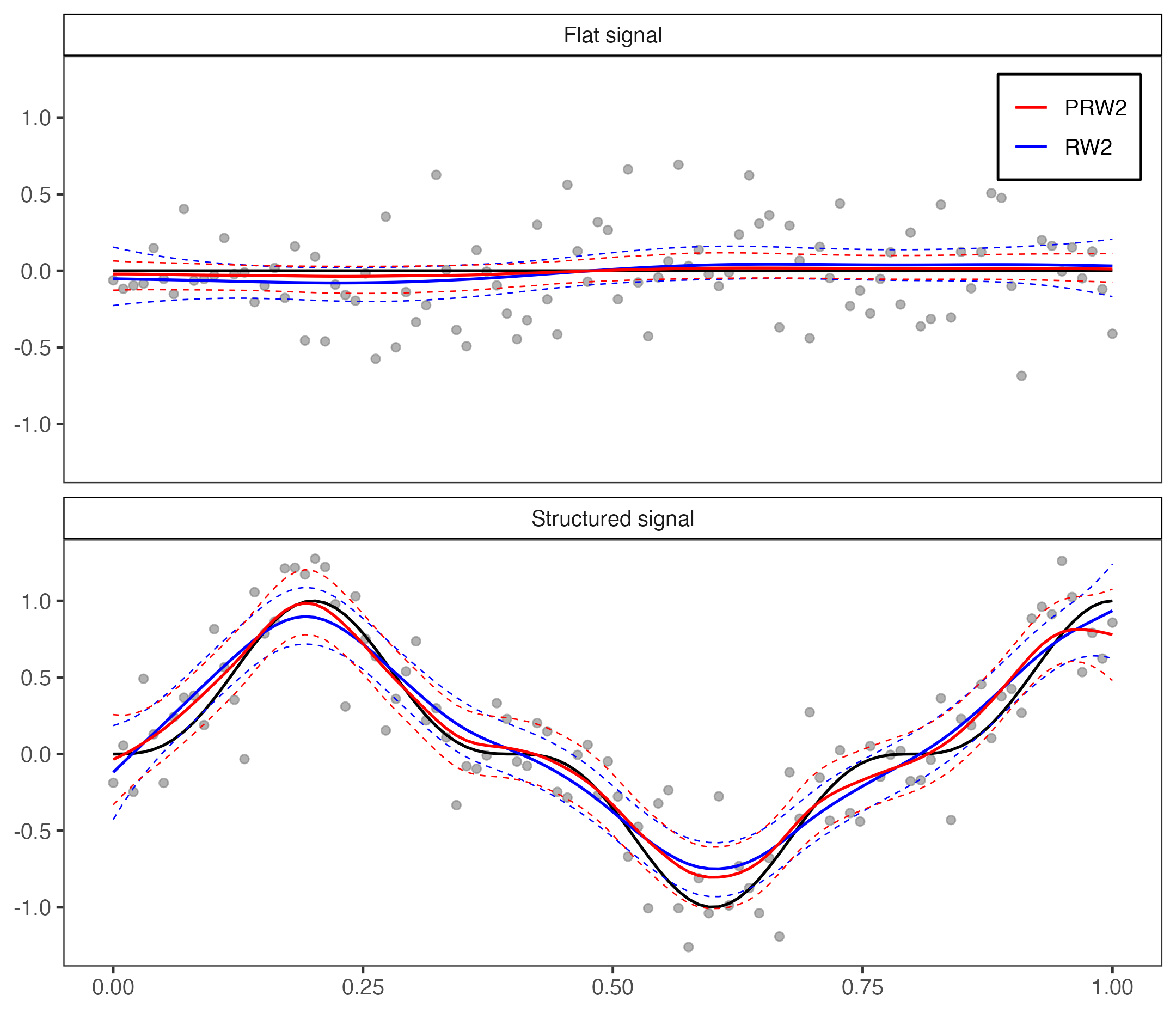}
\caption{Dense/full-data fits of the PRW2 and RW2 for structured and flat signals (black).}
\label{fig:dense-fit}
\end{figure}

\begin{table}[H]
\centering
\scriptsize
\renewcommand{\arraystretch}{1}
\setlength{\tabcolsep}{2pt}
\resizebox{0.95\linewidth}{!}{
\begin{tabular}{llcccc}
\toprule
\rowcolor{green!10}
\textbf{Regime} & \textbf{Signal} &
\multicolumn{1}{c}{\textbf{RMSE (PRW2)}} & 
\multicolumn{1}{c}{\textbf{RMSE (RW2)}} & 
\multicolumn{1}{c}{\textbf{MCI width (PRW2)}} & 
\multicolumn{1}{c}{\textbf{MCI width (RW2)}} \\
\midrule

Dense & Flat       & 0.02 & 0.05 & 0.16 & 0.22 \\
\rowcolor{gray!10}
Dense & Structured & 0.11 & 0.12 & 0.41 & 0.36 \\
Gap   & Flat       & 0.02 & 0.04 & 0.17 & 0.26 \\
\rowcolor{gray!10}
Gap   & Structured & 0.10 & 0.15 & 1.06  & 0.80 \\
Edge  & Flat       & 0.02 & 0.14 & 0.22 & 0.78 \\
\rowcolor{gray!10}
Edge  & Structured & 0.48 & 0.63 & 1.21  & 2.18 \\
\bottomrule
\end{tabular}
}
\caption{Simulation summary: RMSE and mean width of the 95\% pointwise credible intervals (MCI width). For the gap and edge regimes, metrics are computed on the prediction region.}
\label{tab:sim-summary}
\end{table}
With dense observations, the PRW2 and RW2 models produce nearly indistinguishable posterior means for both structured and flat signals (Figure~\ref{fig:dense-fit}), confirming that the PRW2 retains RW2-like smoothing when the data are informative.
Quantitatively, the PRW2 model achieves a smaller RMSE in both signals (flat: $0.02$ vs.\ $0.05$; structured: $0.11$ vs.\ $0.12$), while producing intervals of comparable magnitude (Table~\ref{tab:sim-summary}).

\begin{figure}[H]
\centering
\includegraphics[width=0.6\linewidth]{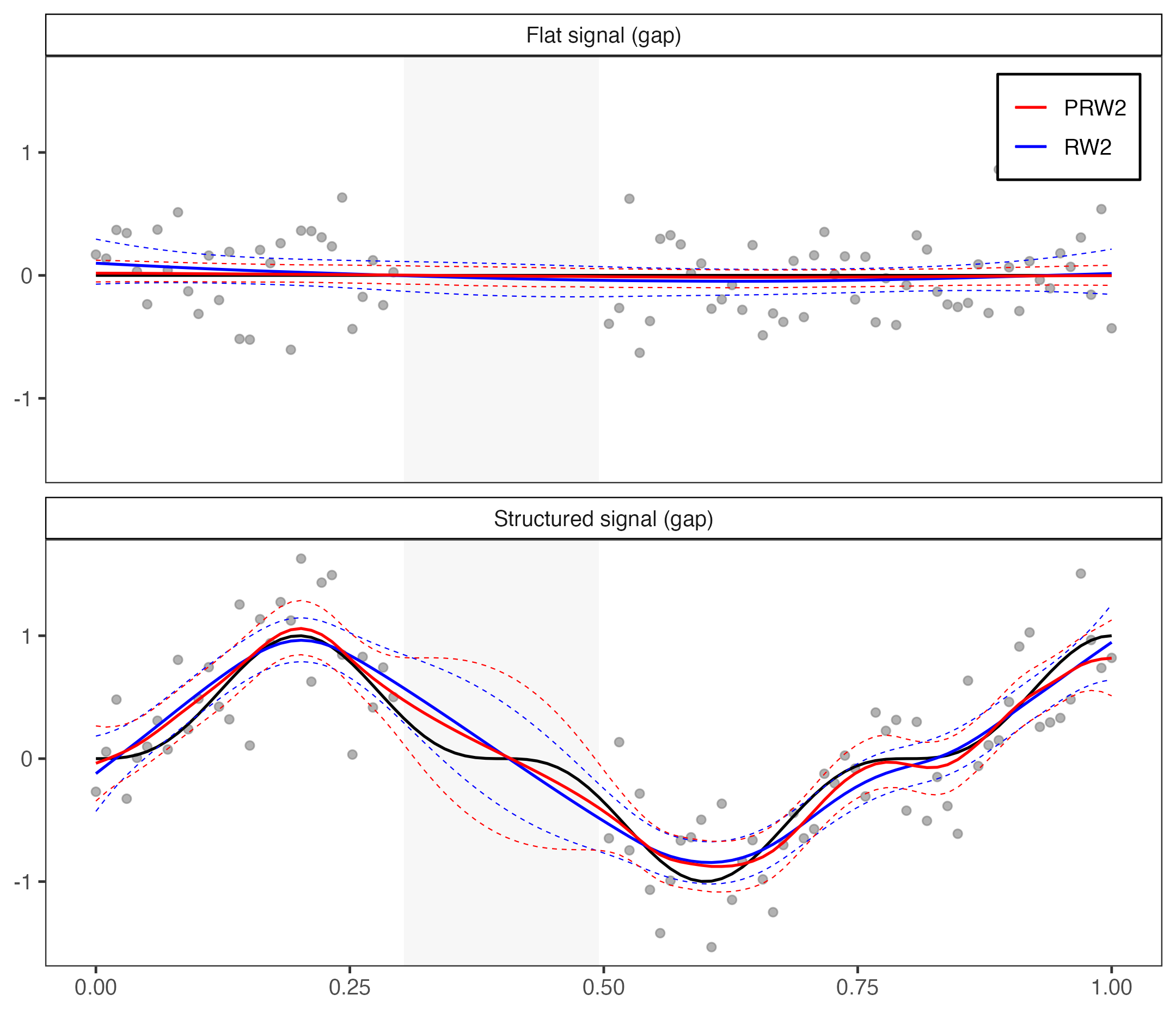}
\caption{Mid-domain interpolation across a data gap by the PRW2 and RW2 models for flat and structured signals (black).}
\label{fig:gap-pred}
\end{figure}

Differences emerge under sparsity. In the flat gap setting, the PRW2 gives tighter uncertainty and lower RMSE (RMSE $0.02$ vs.\ $0.04$; mean CI width $0.17$ vs.\ $0.26$), reflecting its ability to pool information across the domain without inflating uncertainty in a data-deficient interior region (Figure~\ref{fig:gap-pred}).
For the structured gap, the PRW2 achieves a markedly lower RMSE ($0.10$ vs.\ $0.15$) but wider intervals (mean width $1.06$ vs.\ $0.80$).
This widening is concentrated in the gap region (Figure~\ref{fig:gap-pred}), indicating that the PRW2 expresses increased uncertainty precisely where local information is missing, rather than enforcing uniform smoothing.

\begin{figure}[t]
\centering
\includegraphics[width=0.6\linewidth]{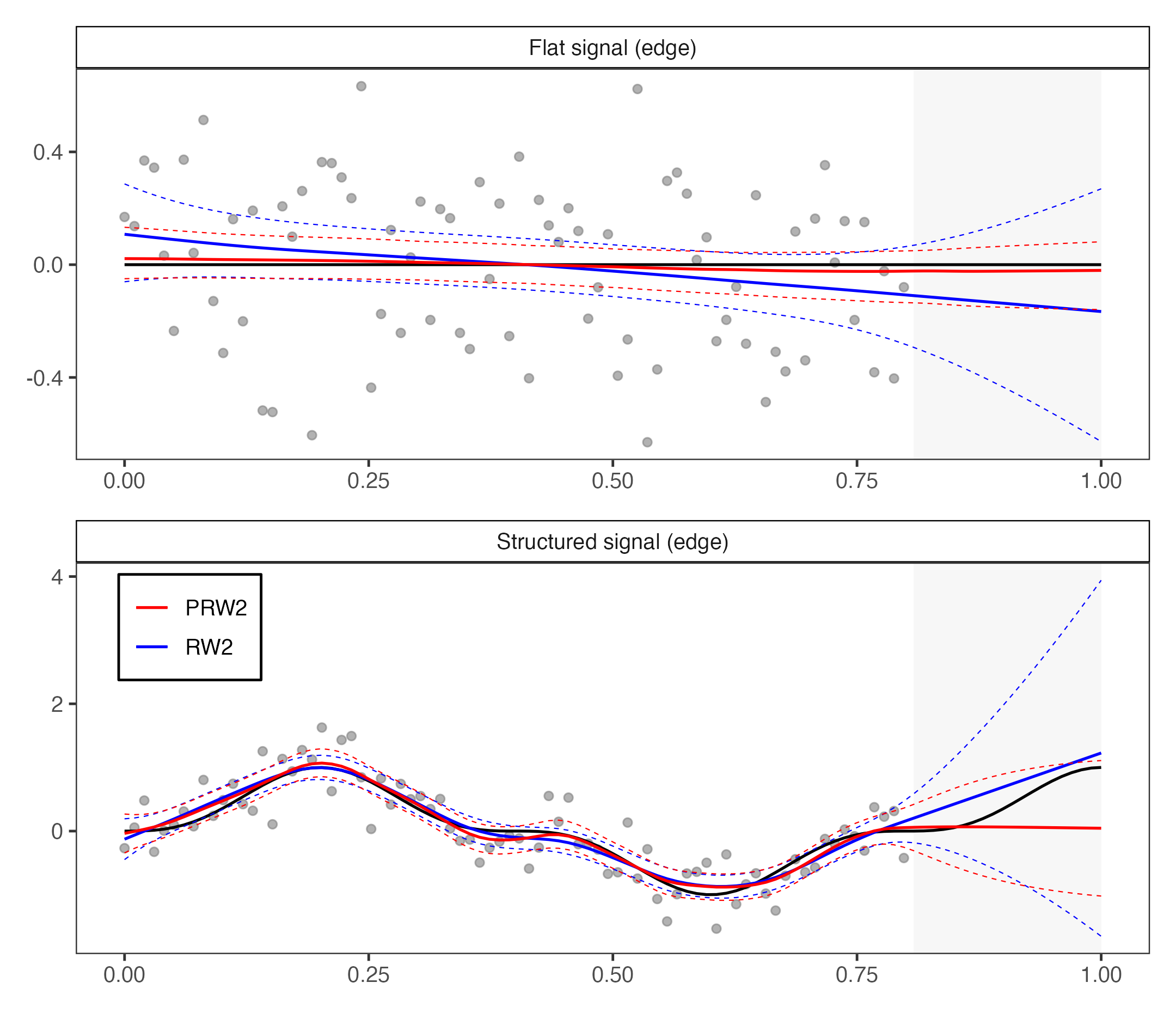}
\caption{The PRW2 and RW2 right-edge prediction/extrapolation for flat and structured signals (black).}
\label{fig:edge-pred}
\end{figure}

At the right boundary, the RW2 exhibits substantial uncertainty inflation, while the PRW2 provides stable extrapolation with moderate credible intervals (Figure~\ref{fig:edge-pred}). In the flat edge setting, this is reflected in a much smaller mean CI width for the PRW2 ($0.22$ vs.\ $0.78$) together with a lower RMSE ($0.02$ vs.\ $0.14$). In the structured edge setting, the intrinsic variance growth leads to very wide intervals (mean width $2.18$), while the PRW2 has less uncertainty (mean width $1.21$) and improves RMSE ($0.48$ vs.\ $0.63$).

Overall, the PRW2 spline matches the RW2 under dense sampling, while providing more stable uncertainty quantification in gaps and near domain boundaries.

\section{Application: Global Youth Mortality Sex Ratios}\label{sec:app}

\subsection{Background and motivation}

Sex-specific mortality estimates for youth aged 15--24 years are central to 
monitoring health disparities and identifying intervention targets. 
\citet{chao2023sex} provided the first global, age-specific assessment of 
mortality sex ratios (MSRs) from 1990 to 2021 in 200 countries and 
territories, documenting a consistent increase in male-to-female mortality with 
age. Their reported global MSR for the 15--24 year group increased from 1.29 (90\% uncertainty interval, UI: 1.23--1.35) in 1990 to 1.65 (90\% UI: 1.52--1.75) in 2021, driven largely by faster declines in female mortality.
Their model used RW1 temporal effects. In such settings, the RW2 can be attractive from the perspective of trend regularization, but its uncertainty behavior may become impractical, especially near boundaries and in sparse regions. Hence, the original study used the RW1 model, the less unstable intrinsic alternative. While this avoids extreme variance inflation, it also reduces the smoothness of the fitted trajectories. This trade-off between stable uncertainty and second-order smoothness motivates the use of the PRW2 model in this reanalysis.

\subsection{Data and model}

We reanalyze the MSR for ages 15--24 in 200 countries and 
territories from 1990 to 2021, using the same harmonized mortality datasets, 
likelihood, and hierarchical structure as \citet{chao2023sex}. Data sources comprise vital registration systems, population censuses, and nationally representative household surveys, including the Demographic and Health Surveys (DHS) and Multiple Indicator Cluster Surveys (MICS), yielding series that range from complete annual coverage to sparse or entirely indirect observations.

The mortality sex ratio for country $c$ and year $t$ is defined as
\[
\mathrm{MSR}_{c,t} = \frac{q^{(m)}_{c,t}}{q^{(f)}_{c,t}},
\]
where $q^{(m)}_{c,t}$ and $q^{(f)}_{c,t}$ denote the age-specific 
probabilities of dying between exact ages 15 and 25 for males and females 
respectively. The observed log sex ratio is modeled as
\[
y_{c,t} = \Psi_{c,t} + \varepsilon_{c,t},
\qquad
\varepsilon_{c,t} \sim \mathcal{N}(0,\, u_{c,t}^2 + \omega^2_{s[c,t]}),
\]
where $u_{c,t}^2$ is the known sampling variance and $\omega^2_{s[c,t]}$ 
is the non-sampling error variance for the data-source type $s$. The latent 
process decomposes as
\[
\Psi_{c,t} = \Theta_{d[c,t]} + \Phi_{c,t},
\]
where $\Theta_{d[c,t]}$ is the global expected log sex ratio conditional on 
the country's total mortality level, and $\Phi_{c,t}$ is a country-specific 
temporal deviation modeled with the RW1, RW2, or PRW2, depending on the 
specification. We assign PC priors to all precision parameters 
\citep{simpson2017penalising}. Inference is performed via INLA 
\citep{rue2009approximate}, and results are reported as posterior means with 
90\% credible intervals. All three models are fitted to the same data with no 
other modifications, isolating the sole effect of the temporal prior.

The figures show two posterior quantities: the country-specific sex ratio 
$\exp(\Psi_{c,t})$ (red curve, with a 90\% credible band) and the global 
expected sex ratio $\exp(\Theta_{d[c,t]})$ (green curve), following the 
display convention of \citet{chao2023sex}.

\begin{figure}[H]
  \centering

  \begin{subfigure}[b]{0.42\textwidth}
    \includegraphics[width=\linewidth]{\detokenize{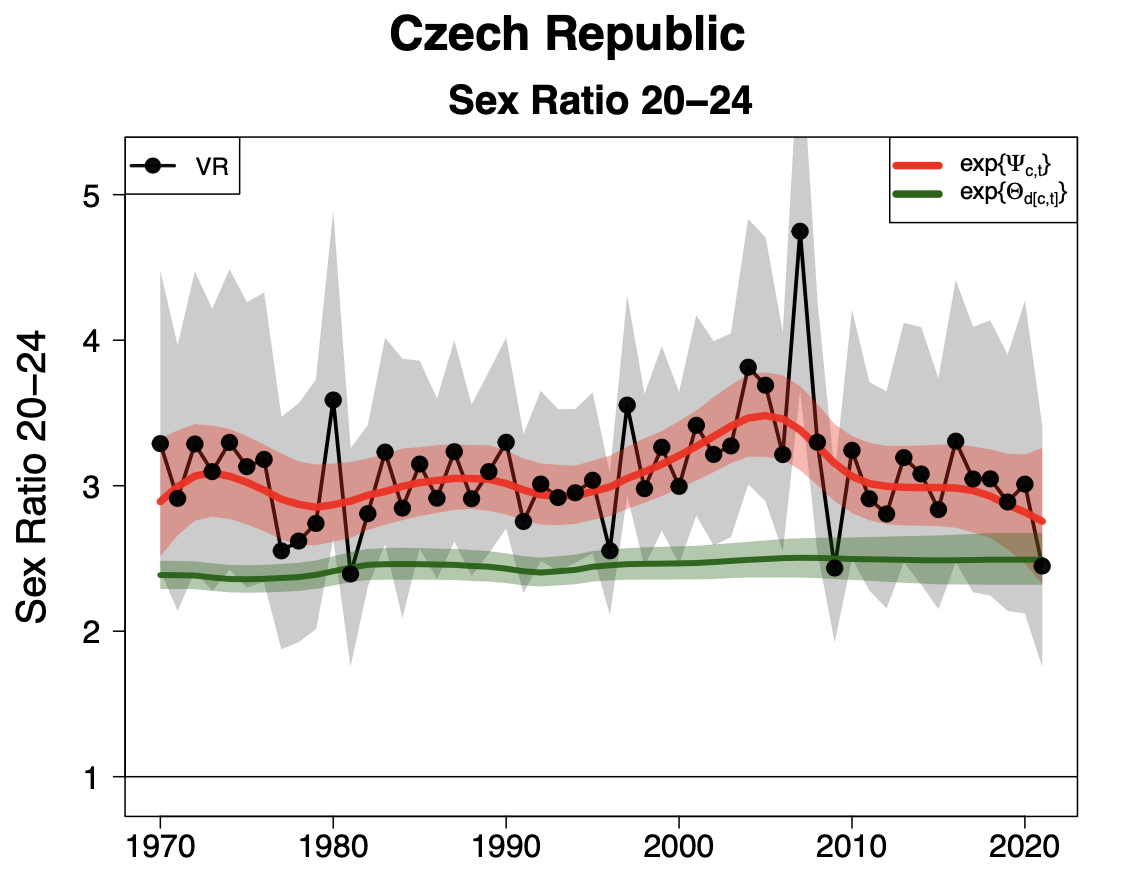}}
  \end{subfigure}\hfill
  \begin{subfigure}[b]{0.42\textwidth}
    \includegraphics[width=\linewidth]{\detokenize{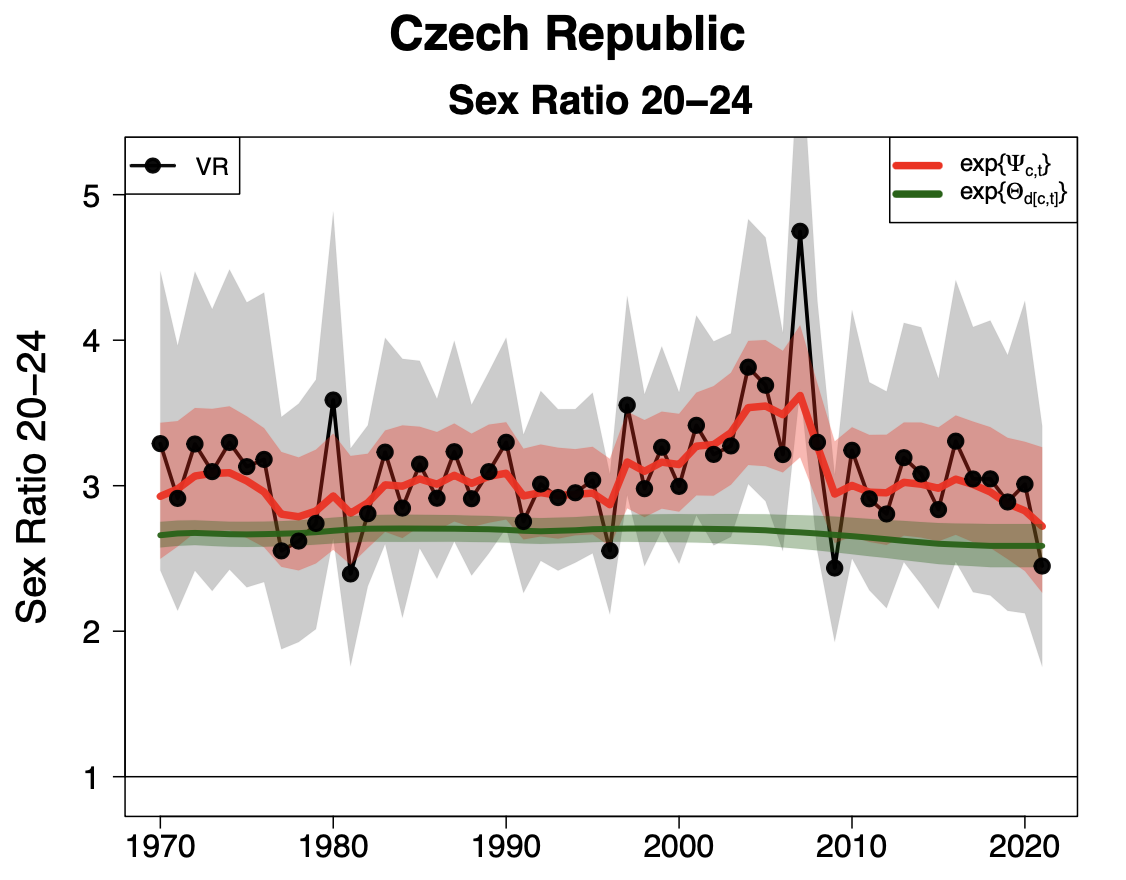}}
  \end{subfigure}

  \vspace{0.3cm}

  \begin{subfigure}[b]{0.42\textwidth}
    \includegraphics[width=\linewidth]{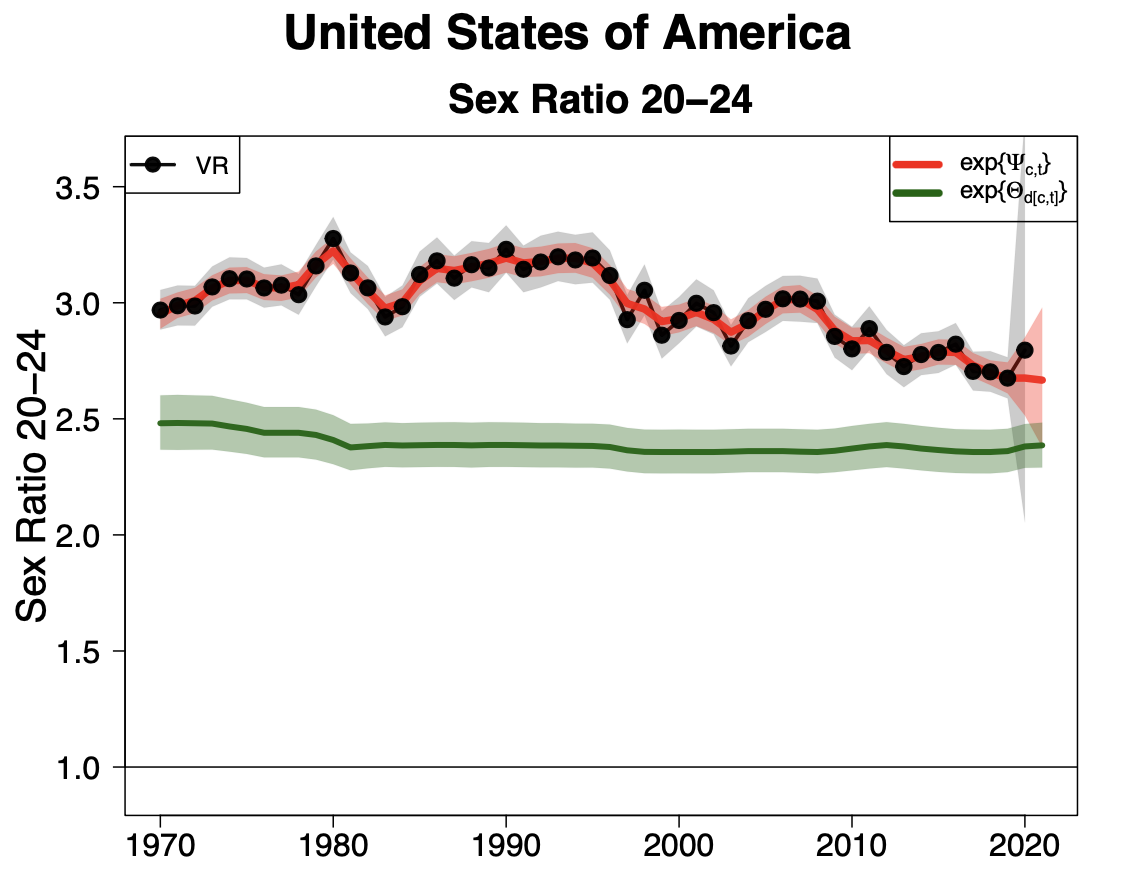}
  \end{subfigure}\hfill
  \begin{subfigure}[b]{0.42\textwidth}
    \includegraphics[width=\linewidth]{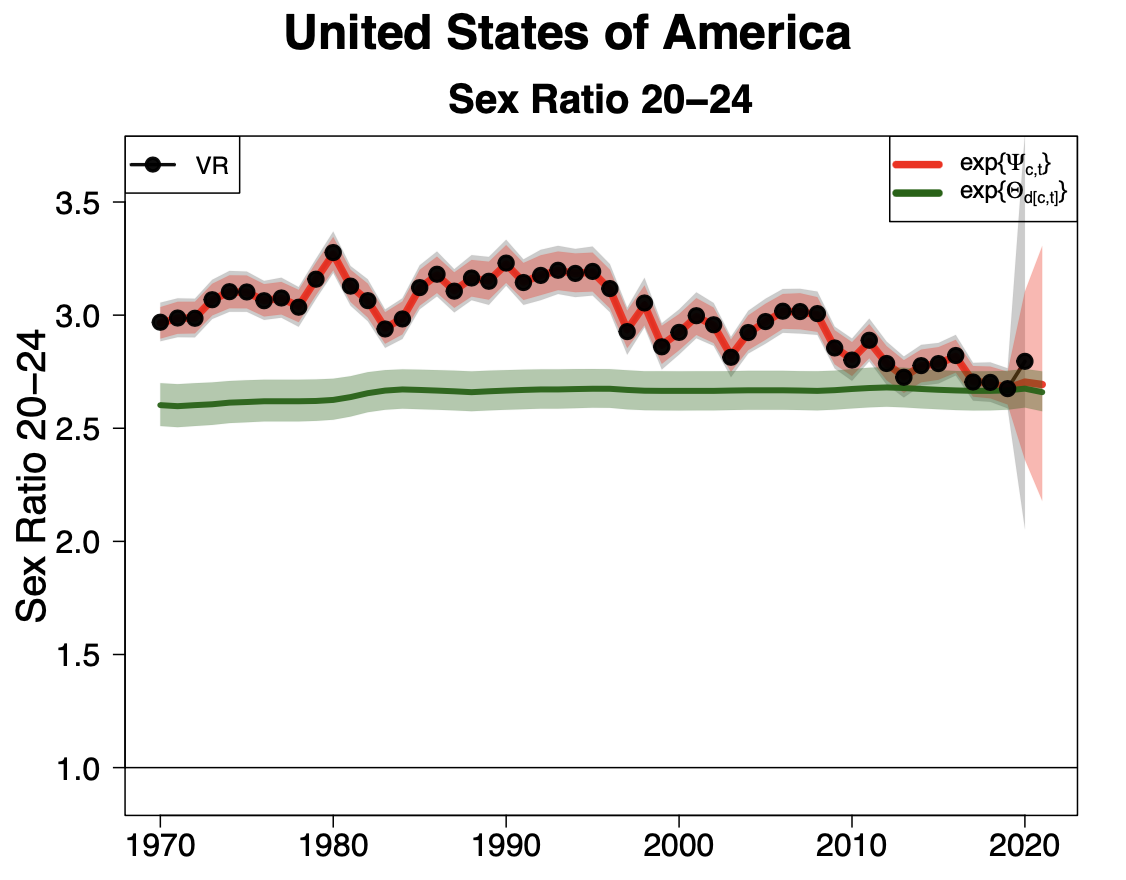}
  \end{subfigure}

  \caption{The PRW2 (left) versus the RW1 (right) in data-rich country series.}
  \label{fig:prw2-vs-rw1-rich}
\end{figure}

\begin{figure}[H]
  \centering

  \begin{subfigure}[b]{0.45\textwidth}
    \includegraphics[width=\linewidth]{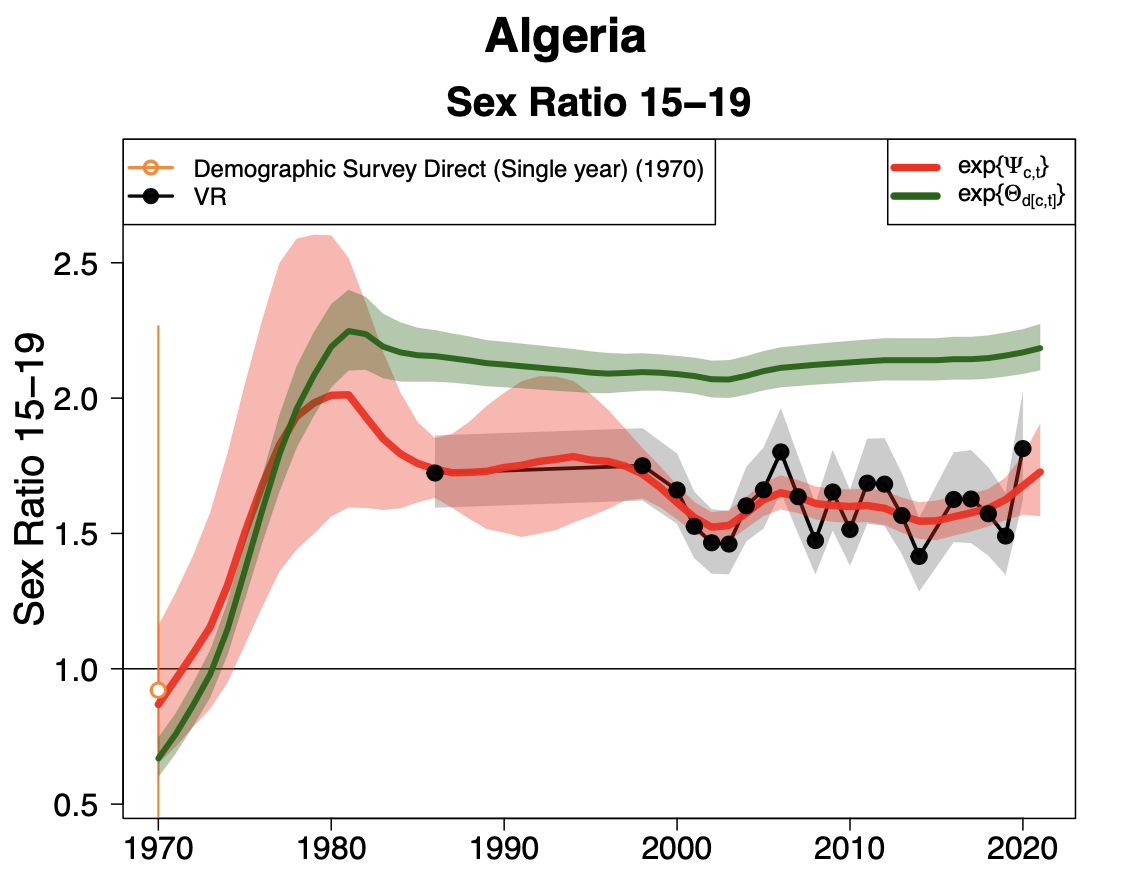}
  \end{subfigure}\hfill
  \begin{subfigure}[b]{0.45\textwidth}
    \includegraphics[width=\linewidth]{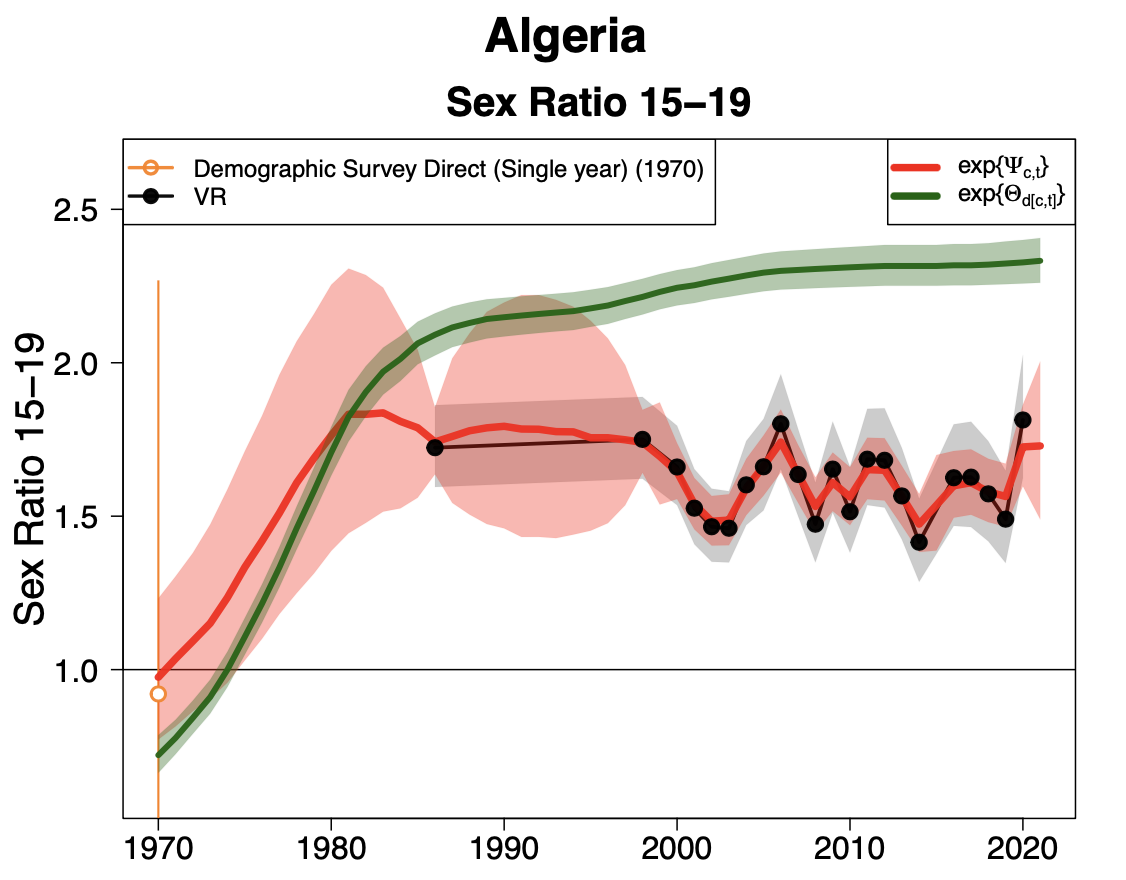}
  \end{subfigure}

  \vspace{0.3cm}

  \begin{subfigure}[b]{0.42\textwidth}
    \includegraphics[width=\linewidth]{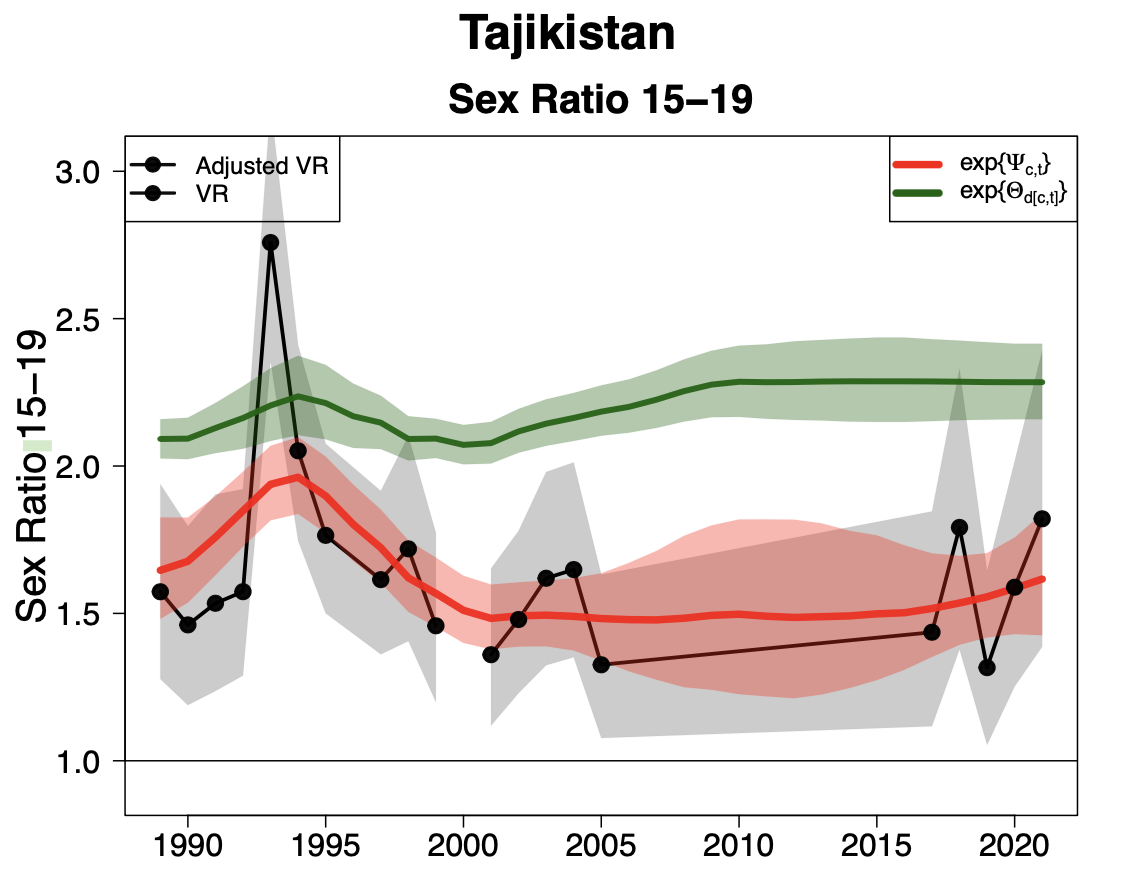}
  \end{subfigure}\hfill
  \begin{subfigure}[b]{0.42\textwidth}
    \includegraphics[width=\linewidth]{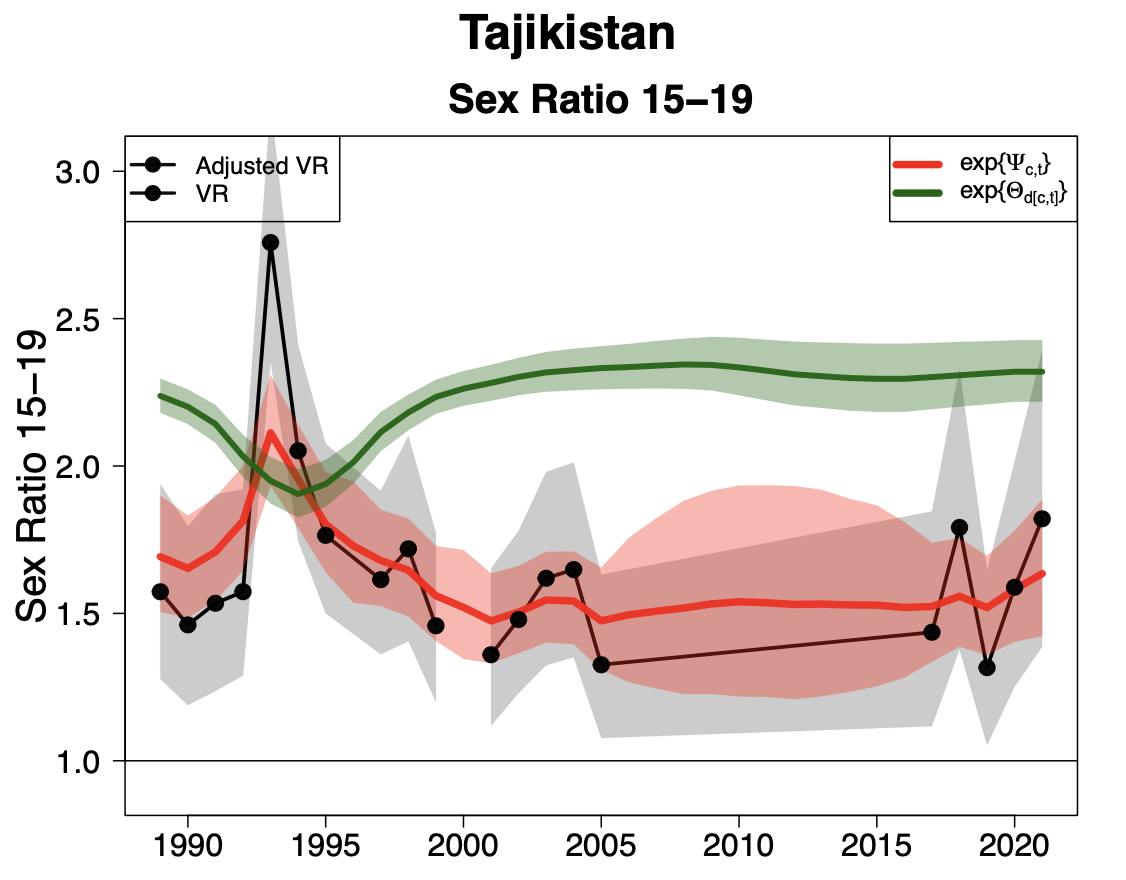}
  \end{subfigure}

  \caption{The PRW2 (left) versus the RW1 (right) in data-sparse country series.}
  \label{fig:prw2-vs-rw1-sparse}
\end{figure}

\subsection{Results}

In countries with sustained observation histories, the PRW2 and RW1 models produce closely aligned posterior means, but the PRW2 produces materially smoother trajectories by suppressing year-to-year volatility that the RW1 model's first-order structure cannot filter (Figure~\ref{fig:prw2-vs-rw1-rich}). In data-sparse settings -- where the intrinsic diffusion of the RW1 model leads to credible intervals widening through observation gaps and at series boundaries -- the PRW2 maintains stable, bounded uncertainty throughout (Figure~\ref{fig:prw2-vs-rw1-sparse}). The contrast is sharpest in countries with intermittent survey coverage, such as Tajikistan (ages 15--19) after the early 2000s: the RW1 model's 90\% band expands rapidly once observations become sparse, while the PRW2 model's band remains narrower.

\begin{figure}[htbp]
  \centering

  \begin{subfigure}[b]{0.45\textwidth}
    \includegraphics[width=\linewidth]{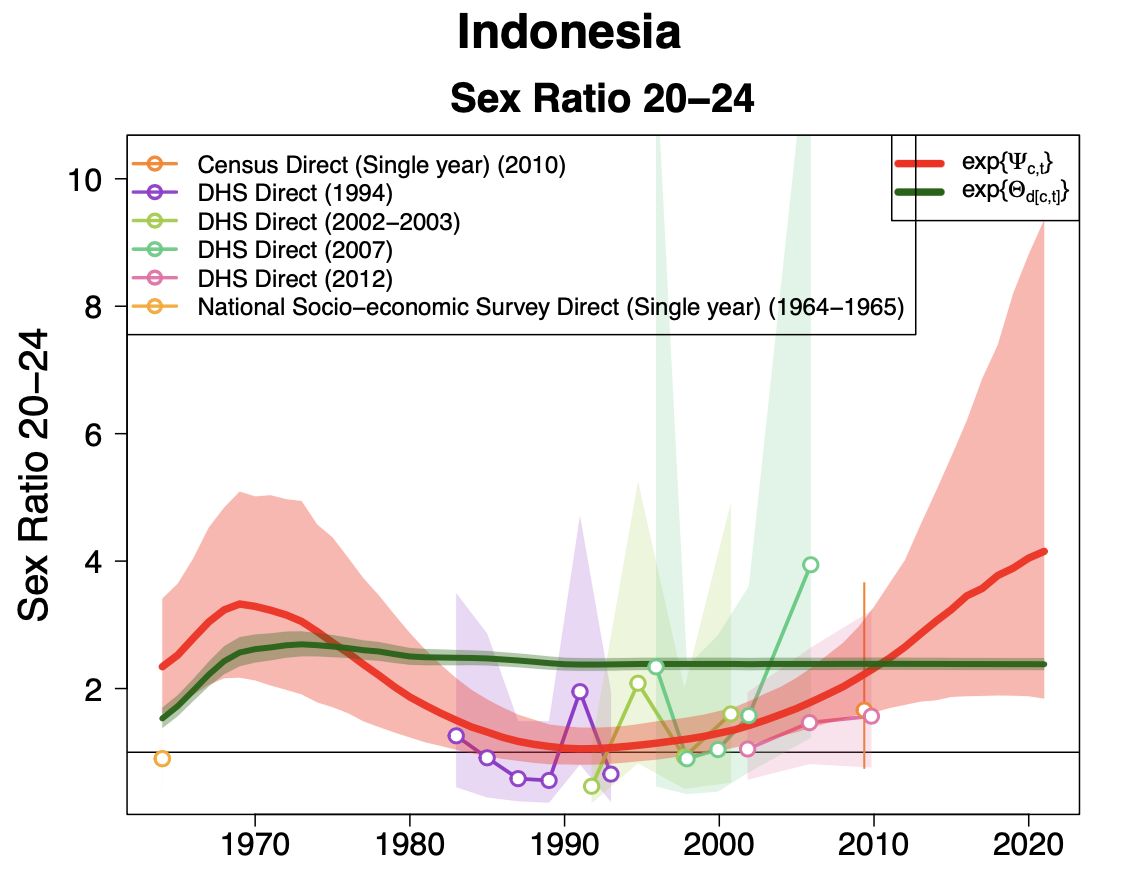}
  \end{subfigure}\hfill
  \begin{subfigure}[b]{0.45\textwidth}
    \includegraphics[width=\linewidth]{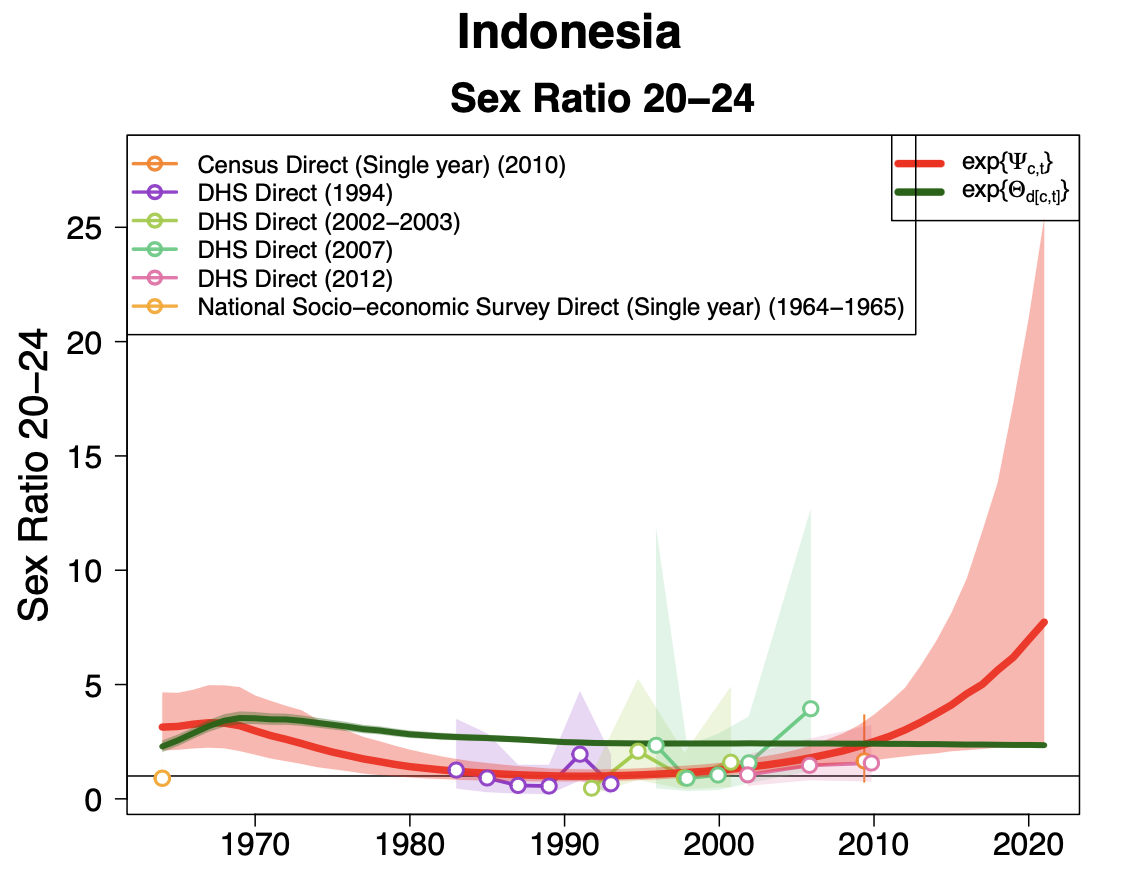}
  \end{subfigure}

  \vspace{0.3cm}

  \begin{subfigure}[b]{0.45\textwidth}
    \includegraphics[width=\linewidth]{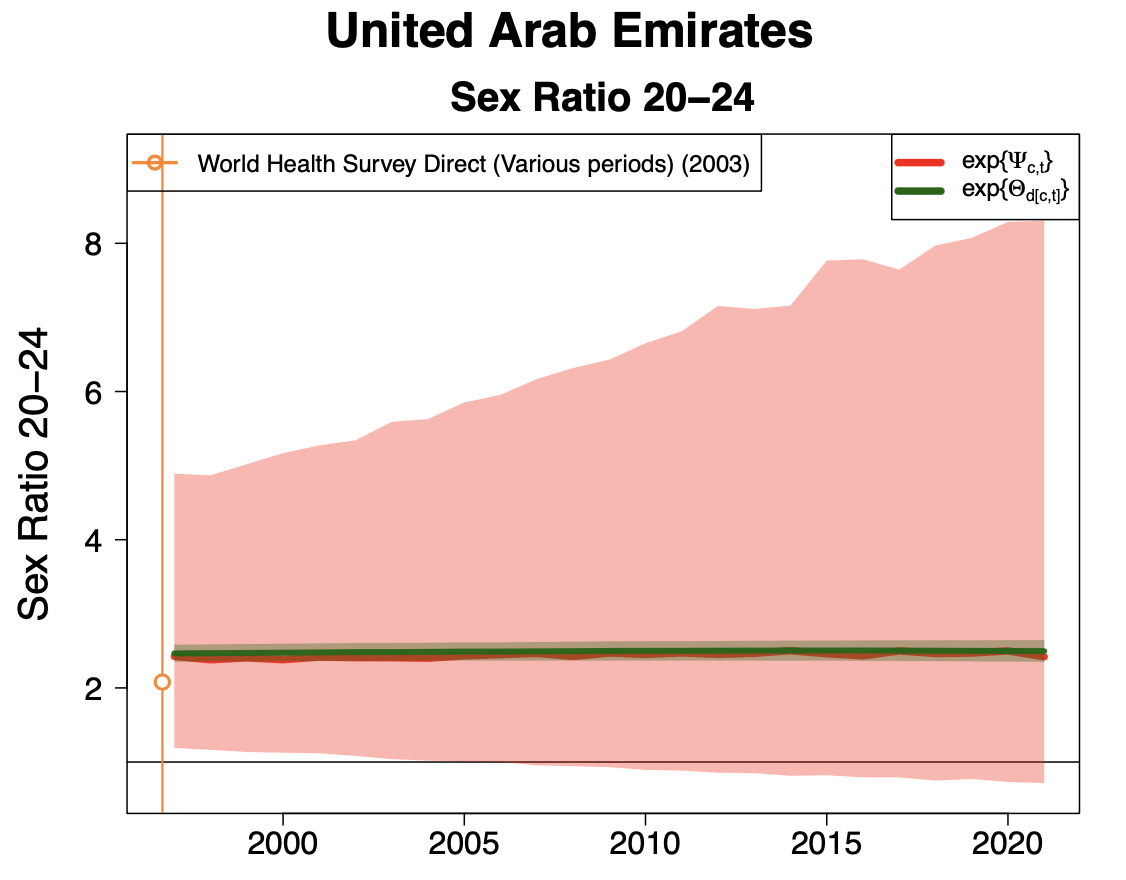}
  \end{subfigure}\hfill
  \begin{subfigure}[b]{0.45\textwidth}
    \includegraphics[width=\linewidth]{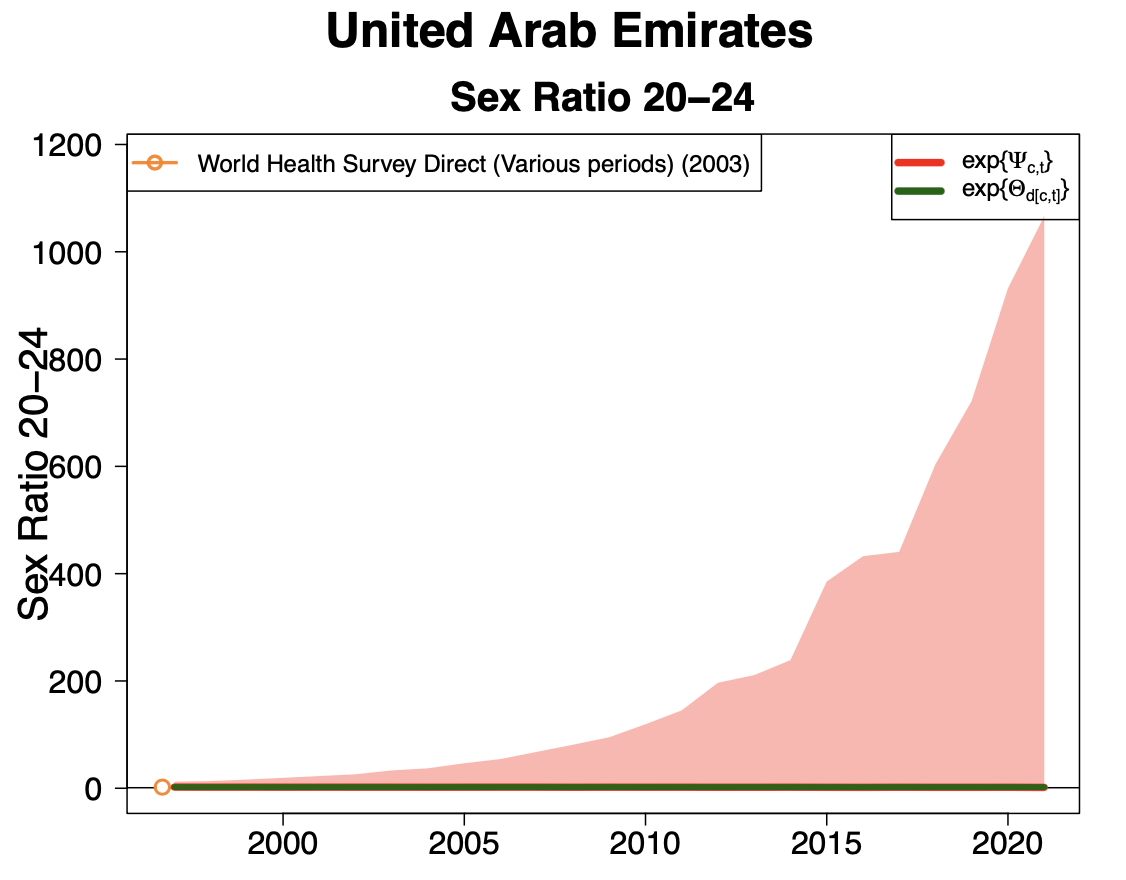}
  \end{subfigure}

  \vspace{0.3cm}

  \begin{subfigure}[b]{0.45\textwidth}
    \includegraphics[width=\linewidth]{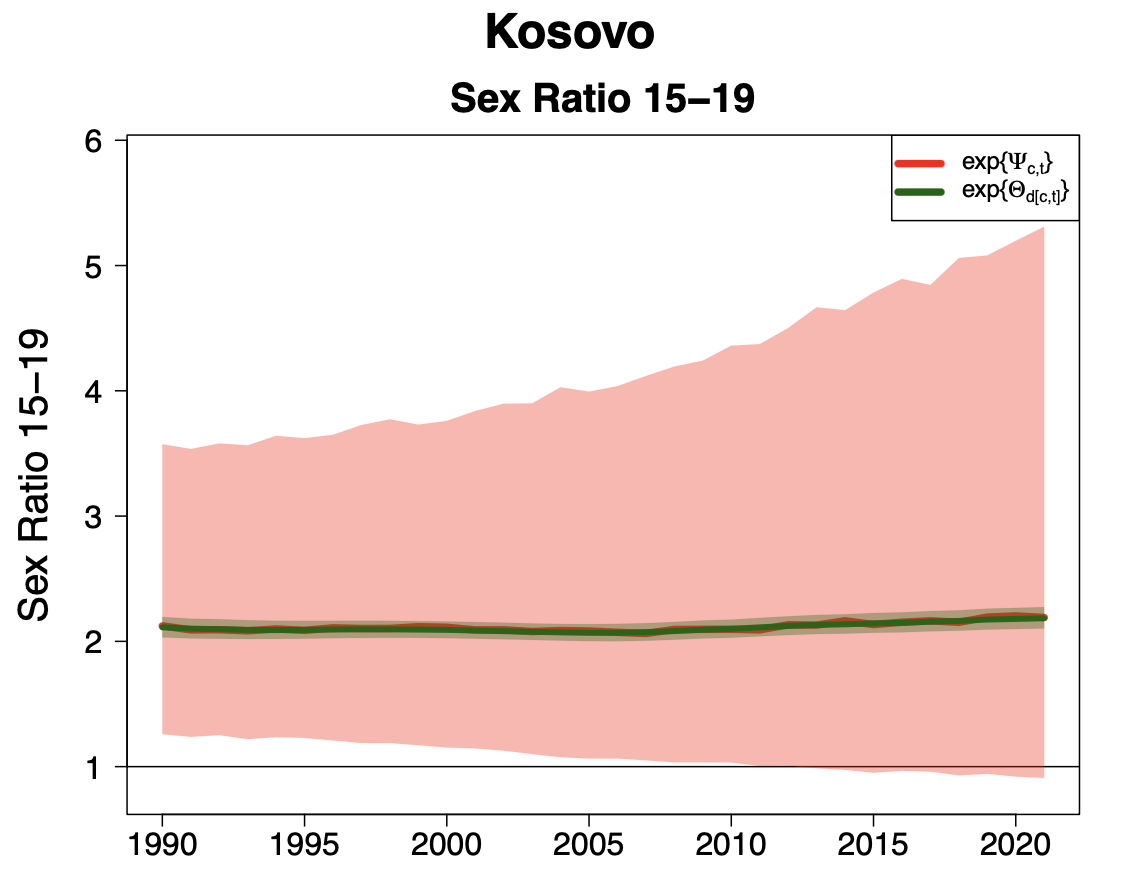}
  \end{subfigure}\hfill
  \begin{subfigure}[b]{0.45\textwidth}
    \includegraphics[width=\linewidth]{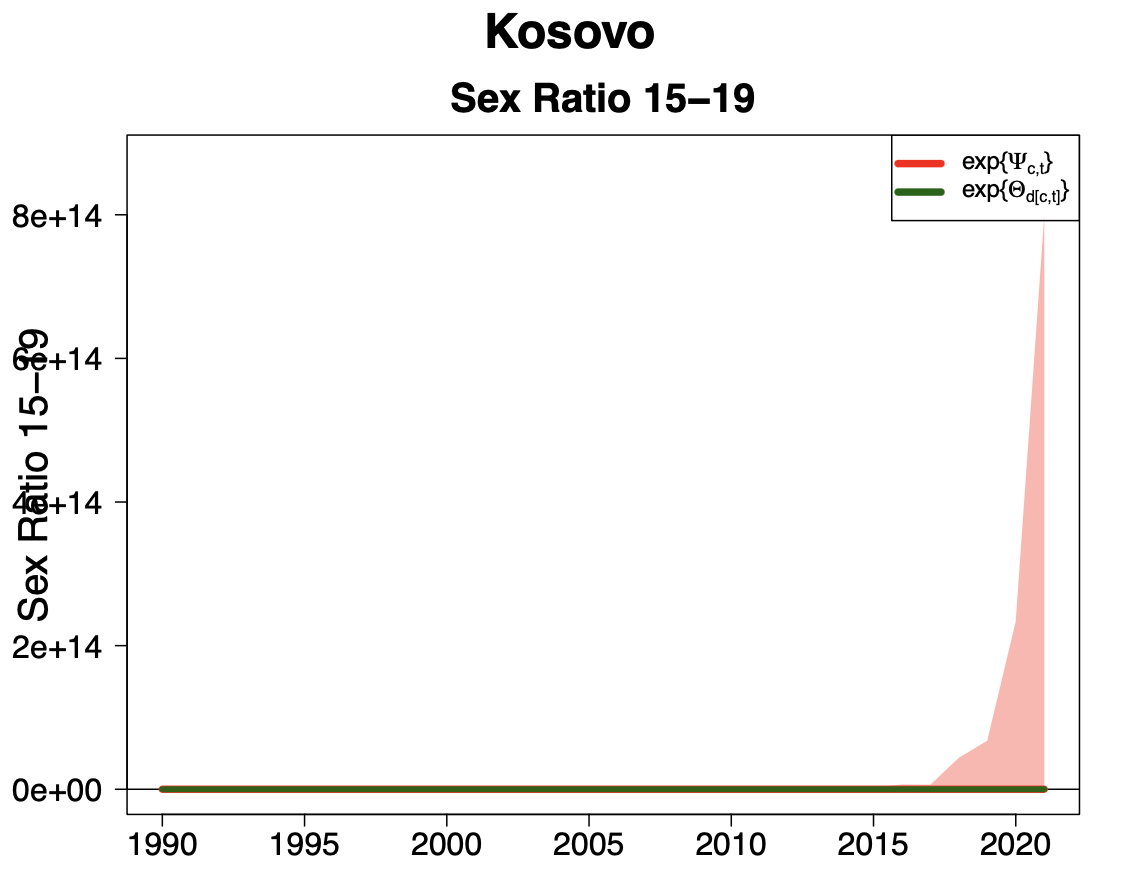}
  \end{subfigure}

  \caption{Extreme uncertainty of the RW2 (right) versus the PRW2 (left) in the
absence of data. The RW2 credible bands expand by orders of magnitude and the
posterior mean drifts from the observations; the PRW2 remains stable throughout.}
  \label{fig:prw2-vs-rw2-sparse}
\end{figure}

Both the RW2 and PRW2 models produce second-order smooth trajectories, but only the PRW2 
avoids the uncertainty expansion that limits the use of the RW2 model in many decision settings. In data-sparse 
countries -- Kosovo (ages 15--19), United Arab Emirates (20--24) and Indonesia (20--24)---the RW2 credible 
bands expand by orders of magnitude wherever observations are absent or 
infrequent (Figure~\ref{fig:prw2-vs-rw2-sparse}, right panels). The distortion is not confined to the uncertainty bands: once the prediction variance explodes, the RW2 
posterior mean itself drifts away from the data, pulled by the prior rather 
than anchored to the observations. 

\begin{figure}[htbp]
  \centering

  \begin{subfigure}[b]{0.45\textwidth}
    \includegraphics[width=\linewidth]{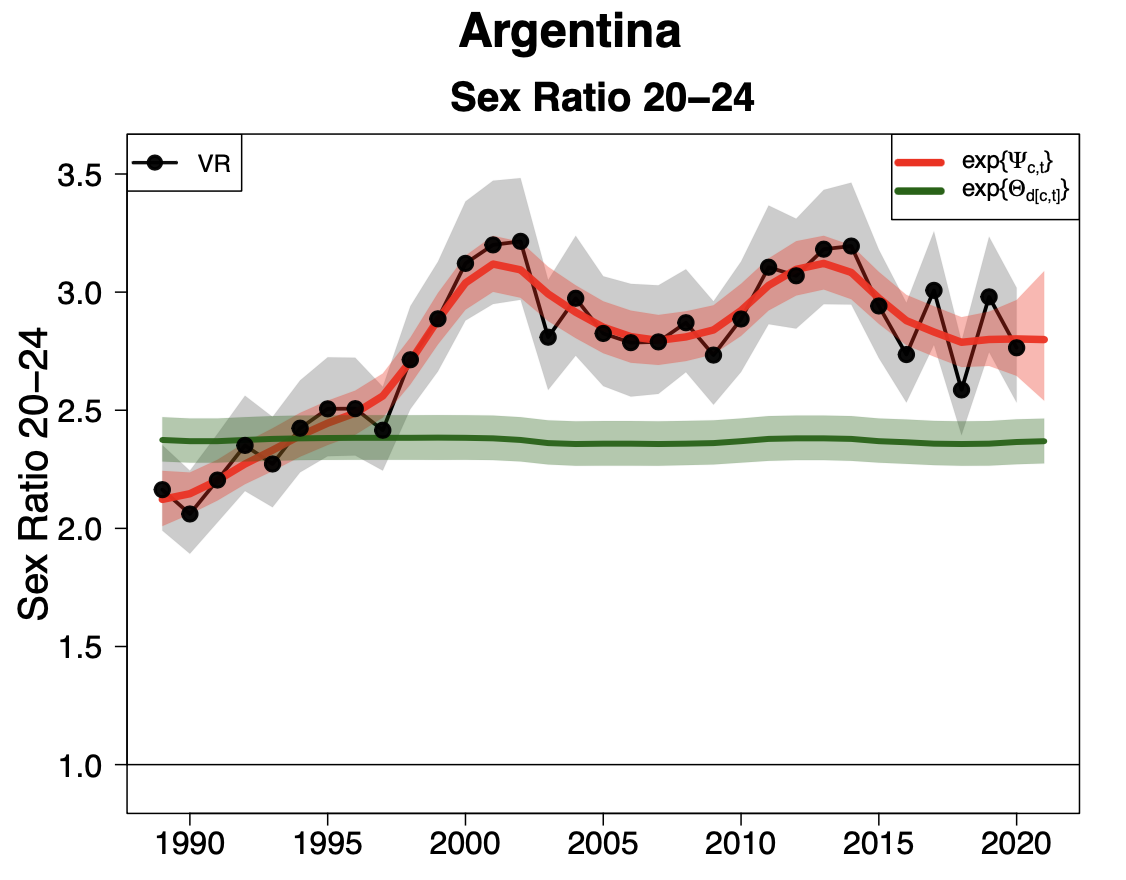}
  \end{subfigure}\hfill
  \begin{subfigure}[b]{0.45\textwidth}
    \includegraphics[width=\linewidth]{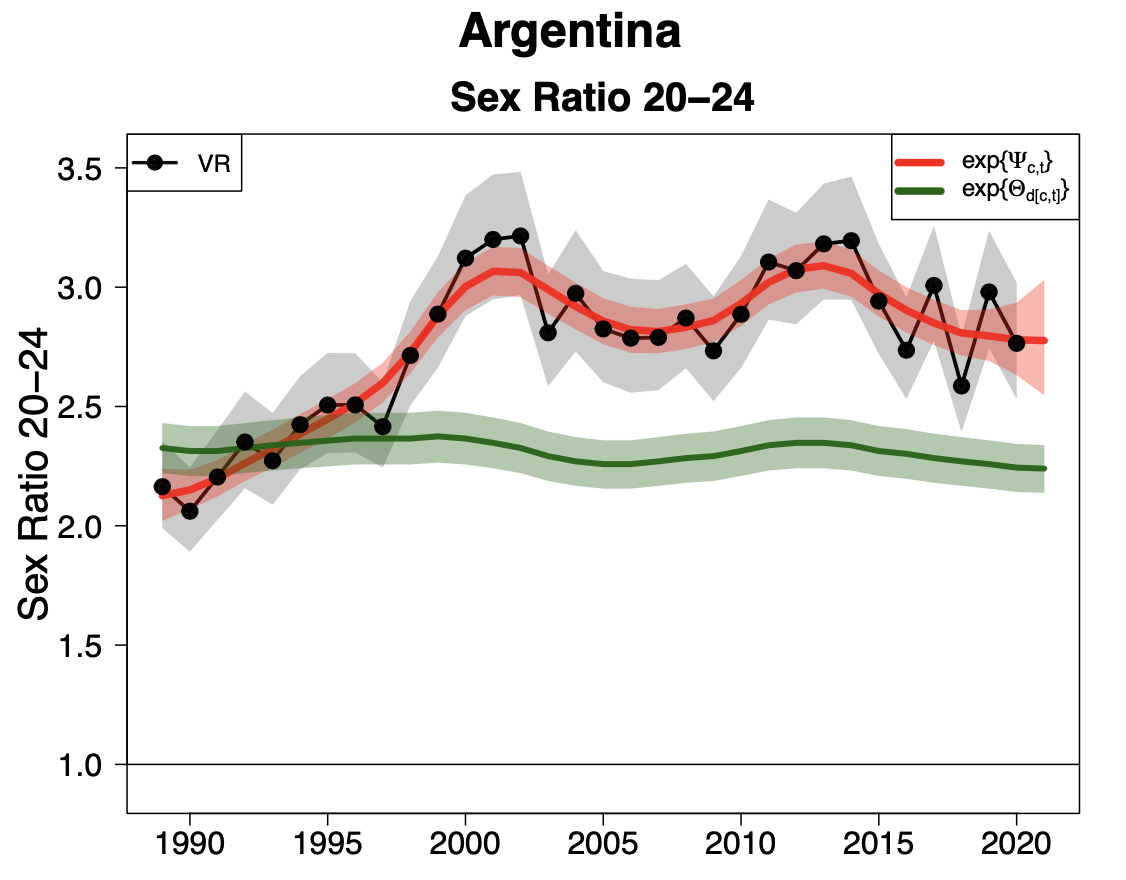}
  \end{subfigure}

  \vspace{0.3cm}

  \begin{subfigure}[b]{0.45\textwidth}
    \includegraphics[width=\linewidth]{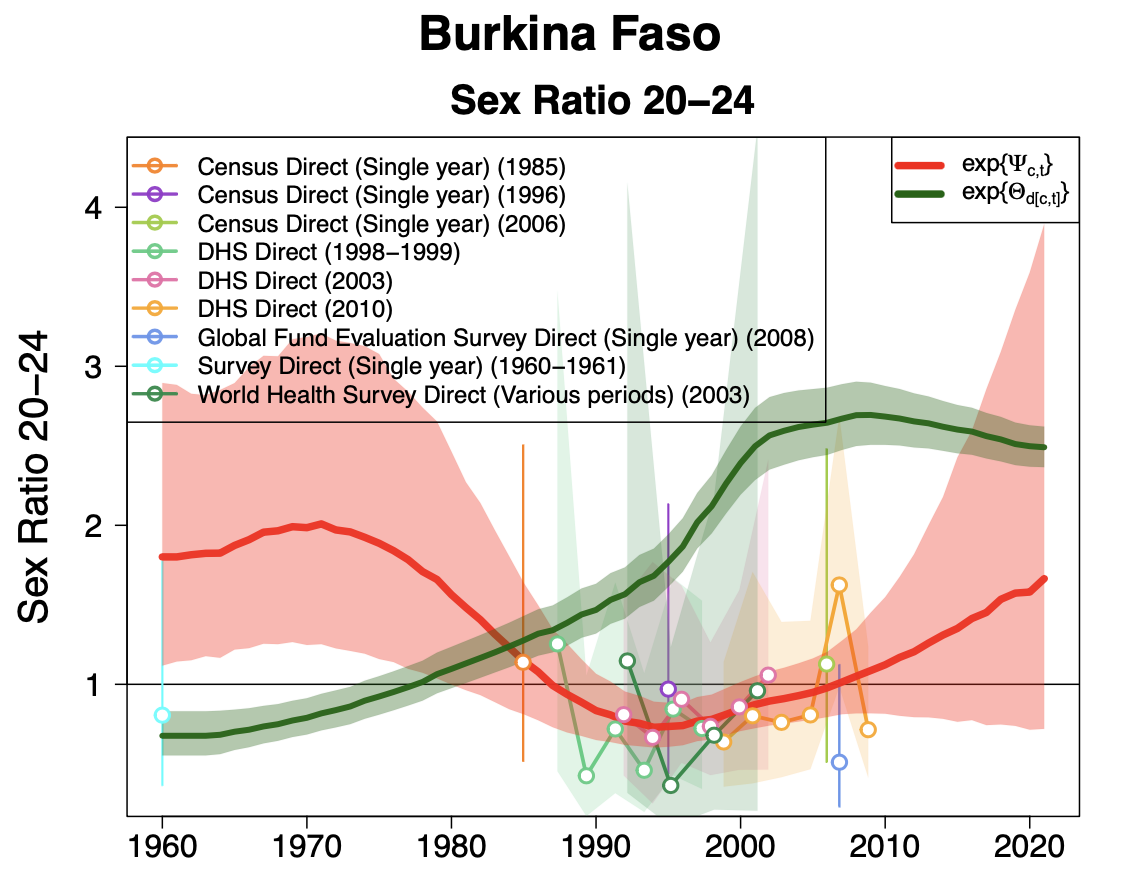}
  \end{subfigure}\hfill
  \begin{subfigure}[b]{0.45\textwidth}
    \includegraphics[width=\linewidth]{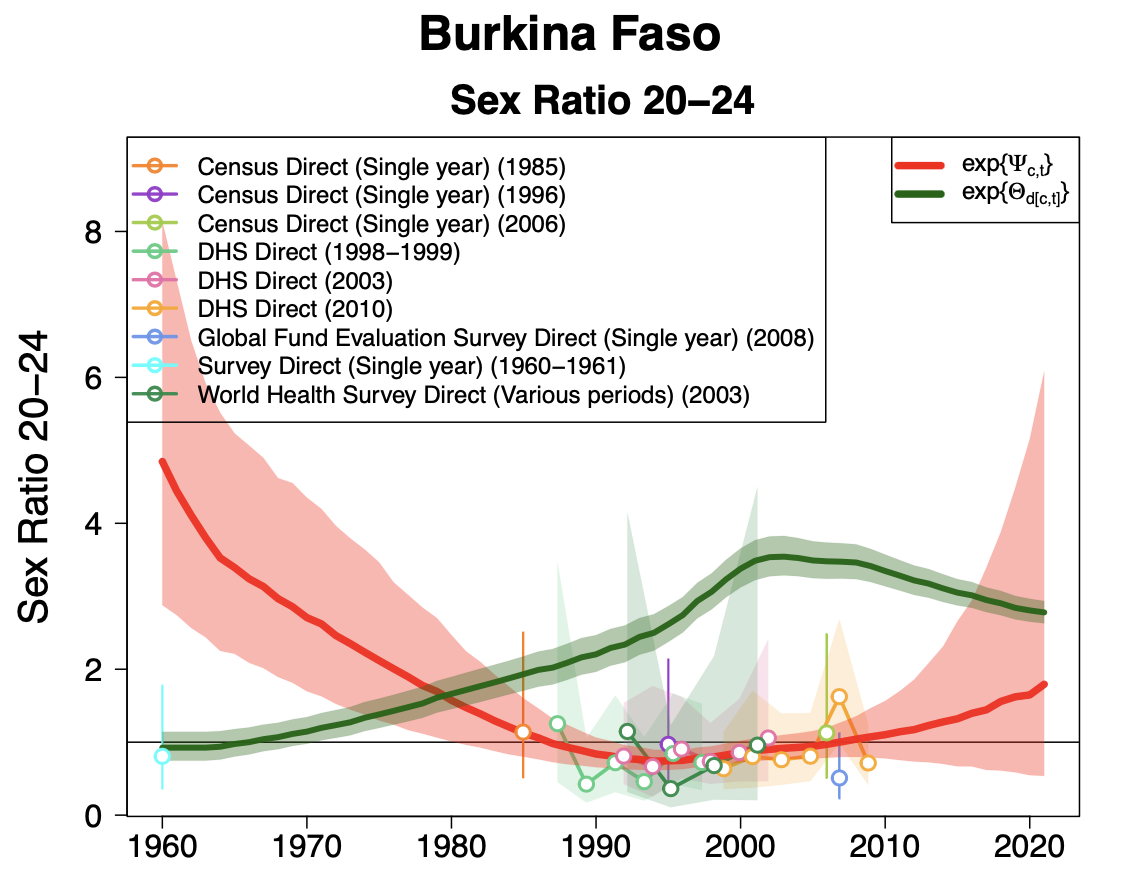}
  \end{subfigure}

  \vspace{0.3cm}

  \begin{subfigure}[b]{0.45\textwidth}
    \includegraphics[width=\linewidth]{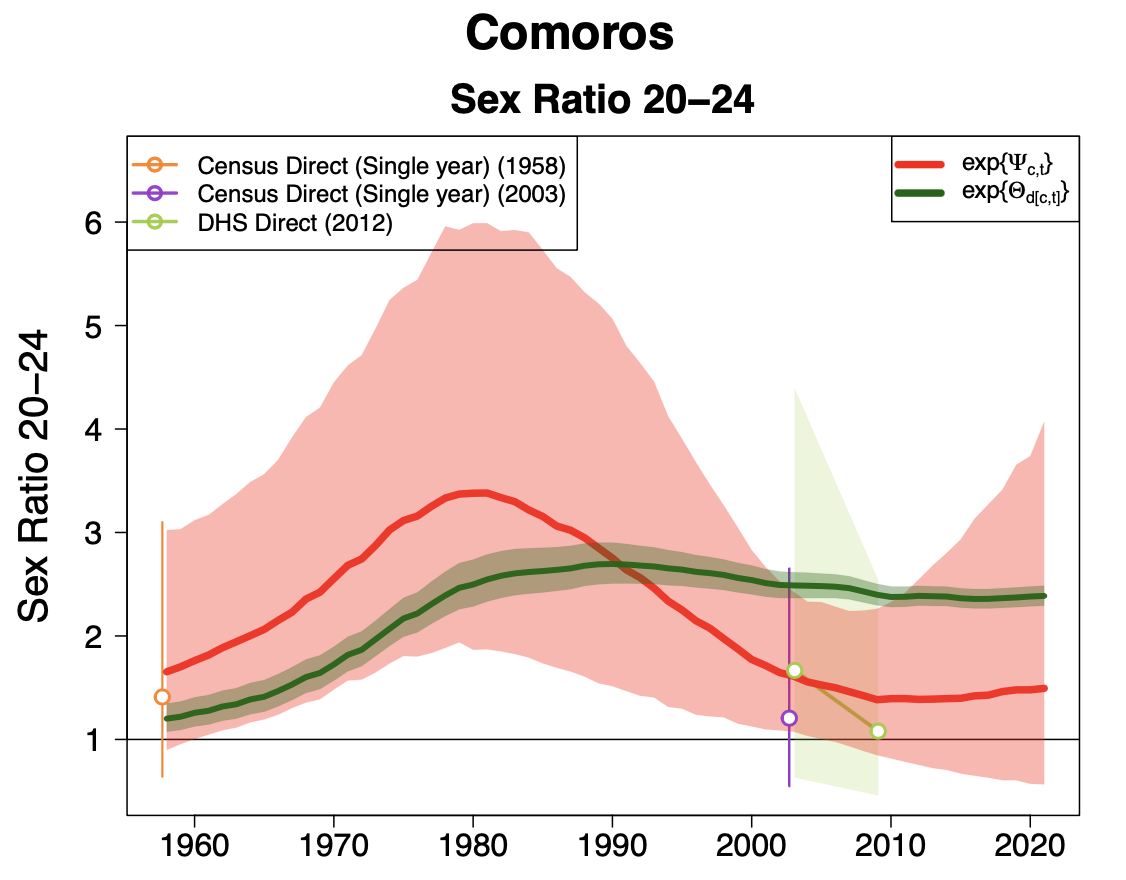}
  \end{subfigure}\hfill
  \begin{subfigure}[b]{0.45\textwidth}
    \includegraphics[width=\linewidth]{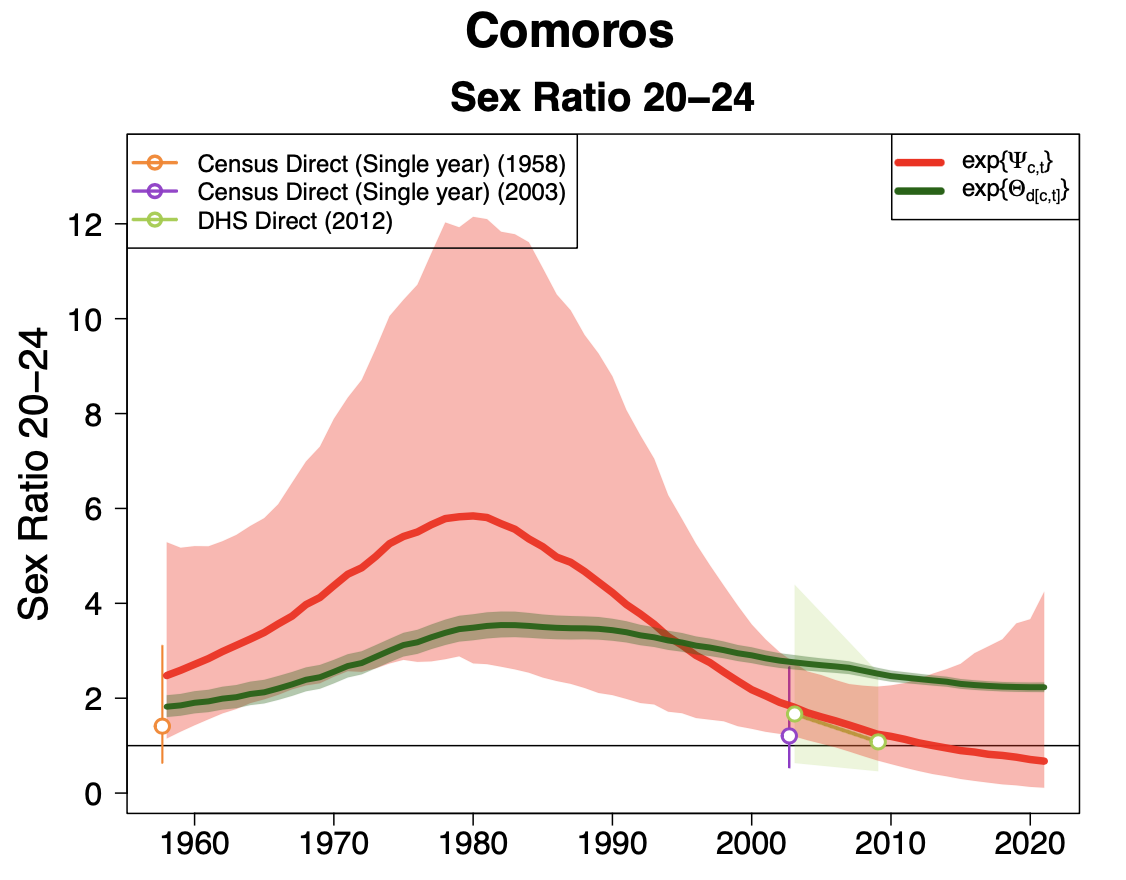}
  \end{subfigure}

  \caption{Uncertainty behaviour of the PRW2 (left) versus the RW2 (right). In
data-rich series (Argentina), the two models agree closely, confirming the PRW2
incurs no smoothing penalty under dense observations. In moderately sparse
series (Burkina Faso, Comoros), the RW2 uncertainty compounds while the PRW2 remains stable.}
\label{fig:prw2-vs-rw2-moderate}
 
\end{figure}

The genuine agreement between the RW2 and PRW2 models is confined to countries with dense 
and uninterrupted observation histories such as Argentina (ages 20--24), where the 
data are sufficiently informative to keep the RW2 stable 
(Figure~\ref{fig:prw2-vs-rw2-moderate}, top panels). In an analysis spanning 200 
countries and age groups, it is impractical to monitor case-by-case whether 
each series provides sufficient data to keep RW2 stable. The PRW2 removes this 
monitoring burden: its structure ensures coherent results 
across the entire analysis by design.
\subsection{Summary}

The PRW2 resolves the central tension in temporal smoothing for global demographic 
analyses: the choice between smoothness and stable uncertainty is no longer 
forced. The RW1 model avoids explosive uncertainty at the cost of under-smoothing; the RW2 model achieves second-order smoothness at the cost of credible intervals that 
are unreliable wherever observations are sparse. The PRW2 delivers second-order smooth 
trajectories with bounded, interpretable credible intervals across the full 
spectrum of country data densities, without any per-series tuning or post hoc 
triage. For large comparative analyses where automated reproducible inference 
across heterogeneous series is essential, the PRW2 provides a principled drop-in 
replacement for both benchmarks.

\FloatBarrier
\section{Discussion}
\label{sec:disc}
We have developed the PRW2 (PRW1) model as a proper, stationary alternative to the RW2 (RW1) model that retains its spline smoothing property and sparse structure while removing the variance pathologies of the intrinsic formulation. The simulation study and the youth-mortality reanalysis bear out this design: the PRW2 behaves like the RW2 where data are informative, and in sparse regions it keeps uncertainty bounded and stable where the RW2 would inflate without bound.\\ 
\indent The variance behaviors that motivate the PRW2 development are not unique to one-dimensional RW2 splines. A close analog is the conditional autoregressive (CAR) model --- in its intrinsic form, the ICAR or Besag \citep{besag1995conditional} model --- used for areal spatial effects, where marginal variances can depend strongly on the graph structure and boundary configuration, producing artifacts in uncertainty that are driven by the neighborhood graph rather than the data \citep{martinez2025homoscedastic}. This broader context reinforces the value of proper, variance-stable alternatives when uncertainty must be comparable across regions, resolutions, or prediction horizons. From this perspective, the PRW2 is a one-dimensional ``properization'' that targets the same class of pathology in spline-based smoothing: it retains the local dependence structure but restores a coherent uniform variance model.\\
\indent From a computational standpoint, the PRW2 retains the key advantage that makes the RW2 appealing in large latent Gaussian models: a sparse/banded precision matrix with linear-time factorization in one dimension. In particular, propriety is achieved without sacrificing sparsity, so posterior computation remains scalable. In practice, this means that the PRW2 can be used as a drop-in replacement for the RW2 in structured additive models when stable uncertainty is required, while maintaining the same asymptotic computational complexity.\\
\indent The Galerkin/FEM construction provides a second benefit beyond interpretability: it offers a principled route to define the same model on non-uniform grids while preserving the inherent Markov structure \citep{lindgren2011explicit}. The accompanying boundary correction is deliberately local: it targets only the boundary block needed to match the stationary joint law at the edges, while leaving the interior conditional model intact. As a result, the discretized field inherits the intended stationary behavior on bounded domains without introducing global constraints that can propagate boundary effects into the interior.\\
\indent Penalized complexity priors provide a transparent mechanism for regularizing the added flexibility \citep{simpson2017penalising}. For the dependence parameter, shrinking toward the infinite-range limit expresses a clear modeling preference: RW2-like behavior is recovered unless the data support a finite-range model. This guards against overfitting by discouraging unnecessarily short ranges, while keeping prior specification interpretable through probability statements on the range scale. \\
\indent Finally, the construction suggests two natural directions for extension. The same ``properization'' principle applies to intrinsic spline models of other orders, yielding non-intrinsic counterparts with bounded uncertainty, and the continuous-time formulation points toward higher-dimensional analogs through the same SPDE/FEM route that yields sparse GMRFs \citep{lindgren2011explicit}. These connections position the PRW2 within a broader strategy: preserve the local Markov structure that enables fast inference, but adopt proper formulations when stable and comparable uncertainty is the primary requirement.

\section*{Data and code availability}
The PRW2 model is available as a built-in latent model in the \texttt{R-INLA}
package (\url{https://www.r-inla.org}), so all analyses can be reproduced using standard inla() calls. The youth‑mortality inputs are those of \citet{chao2023sex}; see that reference for access conditions.

\appendix                   
\input{appendix}

\bibliographystyle{plainnat}
\bibliography{References}

\end{document}

%% file: appendix.tex
%

\section{Marginal variance of intrinsic random-walk models}\label{app:variance}

Consider the intrinsic RW1 and RW2 models on a regular grid of $n$ nodes with unit innovation precision, $\mathbf{Q}_k=\mathbf{D}_k^\top\mathbf{D}_k$, where $\mathbf{D}_k$ is the $k$th-order difference matrix ($k=1,2$). Because $\mathbf{Q}_k$ is rank deficient, marginal variances are defined only under identifiability constraints that remove its null space \citep{rue2005gaussian}: sum-to-zero for RW1, and sum-to-zero together with zero linear trend for RW2. Under these constraints the covariance matrix is the Moore--Penrose generalized inverse $\mathbf{Q}_k^{+}$ \citep{golub2013matrix}. For RW2, a sum-to-zero constraint alone leaves the linear trend unpenalized, and the prior marginal variances are then infinite.

Write $s_i=(i-\tfrac12)/n\in(0,1)$ for the relative location of node $i$. Since $\mathbf{Q}_k$ is $n^{-2k}$ times a discretization of $(-1)^k\,\mathrm{d}^{2k}/\mathrm{d}s^{2k}$ on $[0,1]$, the diagonal of $\mathbf{Q}_k^{+}$ satisfies
\[
\operatorname{Var}(x_i)=n^{2k-1}\,g_k(s_i)\,\{1+o(1)\},\qquad n\to\infty,
\]
where $g_k$ is the diagonal of the generalized Green's function of $(-1)^k\,\mathrm{d}^{2k}/\mathrm{d}s^{2k}$ with the free boundary conditions implied by the intrinsic penalty, restricted to the complement of its null space. Equivalently, $g_k(s)=\operatorname{Var}\{x(s)\}$ where $x$ is the orthogonal projection, onto the complement of $\{1\}$ (RW1) or $\{1,s\}$ (RW2), of Brownian motion (RW1) or integrated Brownian motion (RW2). With $w=s(1-s)$, this gives
\[
g_1(s)=\tfrac13-w,
\qquad
g_2(s)=\tfrac{1}{105}\bigl(1-11w+28w^2+21w^3\bigr).
\]
Both profiles are largest at the boundaries. The RW1 profile is U-shaped and decreases to a quarter of its boundary value at the centre. The RW2 profile is W-shaped: it falls to minima near $s\approx0.21$ and $s\approx0.79$ and rises to a local maximum at $s=\tfrac12$ equal to $21/64$ of its boundary value, so the distortion is not confined to the edges but extends into the interior. Normalizing each profile by its boundary value gives
\[
\tilde g_1(s)=1-3w,\qquad \tilde g_2(s)=1-11w+28w^2+21w^3,
\]
which depend only on relative location and not on $n$. The variance distortion is therefore scale-invariant: it is a fixed function of relative location, so enlarging the domain changes only the overall scale $n^{2k-1}$, not the shape of the distortion (Figure~\ref{fig:variance-sensitivity}). Numerically, $\operatorname{diag}(\mathbf{Q}_k^{+})/n^{2k-1}$ agrees with $g_k(s_i)$ to a maximum relative error of $10^{-6}$ (RW1) and $1.5\times10^{-5}$ (RW2) at $n=1000$.
